\documentclass[prd,aps,preprint,nofootinbib]{revtex4}

\usepackage{amsmath,amssymb,graphicx}
\usepackage{color}

\newcommand{\beq}{\begin{equation}}
\newcommand{\eeq}{\end{equation}}
\newcommand{\bea}{\begin{eqnarray}}
\newcommand{\eea}{\end{eqnarray}}
\def\lla{\big\langle}
\def\rra{\big\rangle}

\def\ssc{\scriptscriptstyle}
\def\lsim{\mathrel{\raise.3ex\hbox{$<$\kern-.75em\lower1ex\hbox{$\sim$}}} }
\def\gsim{\mathrel{\raise.3ex\hbox{$>$\kern-.75em\lower1ex\hbox{$\sim$}}} }

\begin{document}

\preprint{{\vbox{\hbox{NCU-HEP-k063} \hbox{Feb 2016}
}}} \vspace*{.5in}

\title{Analysis on a Nambu--Jona-Lasinio Model of Dynamical Supersymmetry Breaking}

\author{Yifan Cheng }
\email{yifancheng@cc.ncu.edu.tw}

\affiliation{
Department of Physics and 
Center for Mathematics and Theoretical Physics, National Central
University, Chung-Li, Taiwan 32054 }
\author{ Yan-Min Dai }

\affiliation{
Department of Physics and 
Center for Mathematics and Theoretical Physics, National Central
University, Chung-Li, Taiwan 32054 }

\author{ Gaber Faisel }
\email{gfaisel@cc.ncu.edu.tw}

\affiliation{
Department of Physics,
National Taiwan University, Taipei, Taiwan 10617, \\and
Department of Physics, Faculty of  Arts and Sciences,
S\"uleyman Demirel University,  Isparta, Turkey 32260
 }

\author{ Otto C. W. Kong }
\email{otto@phy.ncu.edu.tw}

\affiliation{Department of Physics and 
Center for Mathematics and Theoretical Physics, National Central
University, Chung-Li, Tawian, 32054 }

\begin{abstract}
This is a report  on our newly proposed model of dynamical 
supersymmetry breaking with some details of the analysis involved. 
The model in the simplest version has only a chiral superfield 
(multiplet), with a strong four-superfield interaction in the K\"ahler 
potential that induces a real two-superfield composite with vacuum 
condensate. The latter has supersymmetry breaking parts, which we 
show to bear nontrivial solution following basically a standard 
nonperturbative analysis for a Nambu--Jona-Lasinio type model on a 
superfield setting. The real composite superfield has a spin one 
component but is otherwise quite unconventional. We discuss also 
the parallel analysis for the effective theory with the composite.
Plausible vacuum solutions are illustrated and analyzed. The
supersymmetry breaking solutions have generated soft mass(es) for 
the scalar avoiding the vanishing supertrace condition for the 
squared-masses of the superfield components. We also present some 
analysis of the resulted low energy effective theory with  components 
of the composite become dynamical.  The determinant of the fermionic 
modes is shown to be zero illustrating the presence of  the expected 
Goldstino. The model gives the possibility of constructing a supersymmetric 
standard model with all (super)symmetry breaking masses generated 
dynamically and directly without the necessity of complicated hidden or
mediating sectors.
\end{abstract}

\maketitle

\section{Introduction}
With the discovery of the Higgs particle at the Large Hadron Collider
(LHC), the full success of the Standard Model (SM) has been crowned.
Unfortunately, we still do not see any clear indications of
experimental features beyond so long as phenomenology at the
TeV scale is concerned. Theorists are however mostly unsatisfied with 
the SM, particularly with its Higgs sector and explanation of the origin 
of the electroweak symmetry breaking. With a negative mass-square 
at the right scale put in by hand, the Higgs mechanism looks like only a 
phenomenological description of  the `true' theory behind. Moreover,  
the other parts of the SM theory have their field content tightly 
constrained by the gauge symmetry and no parameters with mass 
dimensions admissible; everything in the Higgs sector looks 
completely arbitrary in comparison. Another way of looking at
the issue would be that the only natural value of any input
mass parameter should be like the model cutoff scale.  We need
a model with a dynamical mechanism to generate the extra
mass scale substantially below the cutoff.

Practical and experimentally accessible physics is really only about 
effective (field) theories. Taking the SM as an effective field theory, one 
would admit the higher dimensional operators with couplings suppressed
by powers of the model cutoff scale in the Lagrangian.  Actually,  a 
dimension six term of four-fermion(/four-quark) interaction with 
otherwise strong coupling gives interesting nonperturbative dynamics 
that can break symmetries and generate masses \cite{C}. That is the 
Nobel prize-winning classic Nambu--Jona-Lasinio (NJL) model \cite{NJL}, 
to which Higgs physics may correspond to the low energy effective
theory with the Higgs doublet being identified as a two-fermion
composite. This beautiful idea of the top-mode SM
\cite{tSM,tSM1,tSM2,tSM3,tSM4,tSM5} fails to accommodate the 
too small phenomenological top quark mass \cite{C}. At this point, 
it looks like a holomorphic supersymmetric version that gives the
(minimal) supersymmetric standard model (SSM) with both Higgs
supermultiplets as two-superfield composite maintains
phenomenological viability \cite{034}.

The SSM is still the most popular candidate theory beyond the
SM being matched to the LHC results. The theoretical beauty of
supersymmetry is certain part of its appeal. The first
supersymmetric Nambu--Jona-Lasinio (SNJL) model was
introduced in the early eighties \cite{BL, BE}, generalizing
the four-fermion interaction to a four-superfield interaction
of the same dimension in the K\"ahler potential . Recently,
our group introduced  the holomorphic version (HSNJL) as an
alternative supersymmetrization \cite{035} with a four-superfield
interaction in the superpotential \cite{034}. The two versions
have different theoretical and phenomenological merits
\cite{034,042,050}. However, both versions require input
soft supersymmetry breaking masses to have the dynamical
(electroweak) symmetry breaking. On the other hand, the
phenomenological SSM requires soft supersymmetry breaking
masses/parameters the origin of which is typically depicted
through elaborated constructions of complicated and contrived
models with extra supersymmetry breaking and mediating
sectors \cite{soft}. Under the background, it is the wish of us
to find a simple model  to get the supersymmetry breaking
and soft mass generation dynamically, hopefully under a
similar framework. That is essentially achieved. We just
reported our first results of a new type of supersymmetric
NJL model with a real two-superfield composite containing
a spin one component. Following and extending the
framework of our earlier analyses \cite{042,050}, we have
established that the model has the gap equation of the
standard real soft mass parameter of the chiral superfield
bearing nontrivial, hence supersymmetry breaking, solution
when the four-superfield coupling is strong enough. The short 
letter we presented the results \cite{062} only gives a sketch 
of the analyses involved and leaves the possibility of a more 
general supersymmetry breaking scenario not fully addressed. 
The current paper is to give a full account of all that.

In Sec. II, we present the model and the supergraph derivation 
of the superfield gap equation, elaborating carefully the 
extension of our framework of analysis \cite{042,050} with 
model parameters and correlation functions taken as superspace 
parameters, like constant superfields, containing supersymmetric 
and supersymmetry breaking parts. The superfield gap equation 
contains components which include wavefunction renormalization
factor and two different soft mass parameters. In Sec.~III, we 
discuss the effective theory picture with the composite and the 
matching effective potential analysis performed at the component 
field level, further strengthen the result and illustrate the physics 
involved. Sec.~IV is devoted to analysis of the nontrivial, 
supersymmetry breaking solutions. In Sec.~V, we go further to 
look at some dynamical features of the composite superfield or 
its various components at low energy, focusing on the Goldstino 
mode. Sec.~VI is devoted to some further discussion of the 
supersymmetry breaking physics. Some remarks and conclusions will 
be presented in the last section. Two appendices are given, the first on 
some details {of} analytical expressions as background for the effective 
theory analysis and some results for two-point functions of the various 
components of the composite superfield relevant for their low energy 
dynamics, and the second on propagator expressions for a (chiral) 
superfield and components admitting the most general mass parameters. 
The latter expressions have not been explicitly presented in the literature.

\section{The model and the superfield gap equation}
The model has a dimension six four-superfield interaction similar but
somewhat different from that of the SNJL model \cite{BL,BE,062}. For the
simplest example, we start with the single chiral superfield (multiplet)
Lagrangian
\footnote{Our basic notation is in line with that of Wess and Bagger \cite{WB}.}
\beq \label{L}
{\cal{L}}=
\int  d^4 \theta \, \left[
\Phi^\dagger \Phi  \, +\frac{m_o}{2}   \Phi \Phi \delta^2(\bar{\theta})
+ \frac{m_o^*}{2} \Phi^{\dagger} \Phi^{\dagger} \delta^2(\theta)
-  \frac{g_o^2 }{2}   \left( \Phi^\dagger \Phi \right)^2 \right] \;,
\eeq
in which we have suppressed any multiplet (color) indices. We illustrate
here a standard NJL gap equation analysis \cite{BL,042,050} applied
to the soft supersymmetry breaking mass parameters, the brief result of
which is reported in Ref.\cite{062}. We are mostly interested in the
generation of the usual soft supersymmetry breaking mass $\tilde{m}^2$
for the superfield $\Phi$. Naively, if the bisuperfield condensate
$\lla \Phi^\dagger \Phi |_{\!\ssc D} \rra$ develops, we would have a
soft supersymmetry breaking mass
$g_o^2 \lla \Phi^\dagger \Phi |_{\!\ssc D} \rra$.  That is where
our key interest is in.

Let us go onto a superfield gap equation analysis following
and extending our earlier formulated framework \cite{042,050}.
To implement an NJL-type gap equation analysis for the
supersymmetry breaking,  the first step of the self-consistent
Hartree approximation is to add the interested soft mass term
$-\int d^4 \theta \Phi^\dagger \Phi \tilde{m}^2 \theta^2 \bar{\theta}^2$
to the free field part and re-subtract it as a mass-insertion type
interaction. The formal gap equation is then given by
\beq \label{gap}
\tilde{m}^2 =
\left.\Sigma_{\tilde{m}}^{\tiny(loop)}(p)\right|_{\mbox{\tiny on-shell}} \;,
\eeq
where $\Sigma_{\tilde{m}}(p)$ is the two-point proper vertex for the 
scalar component $A$ of $\Phi$, as shown in the Fig.~1. Note that the
four-superfield interaction, after the $d^4 \theta$ integration, has the 
part $-g_o^2  A A^\dag (\Phi\Phi^\dag)|_{\theta,\bar{\theta}=0}$. We have 
also performed  the calculation fully in the component field framework 
for case of Ref.\cite{062}, but prefer to illustrate the superfield calculation 
here in accordance with the formulation under the perspective discussed
in Ref.\cite{042}. We consider a superfield two-point proper vertex
$\Sigma_{\Phi\Phi^\dagger}(p;\theta^2, \bar{\theta}^2)$ as taking
value like a constant superfield with components explicitly dependent on
$\theta^2$ and $\bar{\theta}^2$. The $\Sigma_{\tilde{m}}(p)$ of interest
is then to be taken essentially as the $\theta^2\bar{\theta}^2$ component 
of $\Sigma_{\Phi\Phi^\dagger}(p;\theta^2, \bar{\theta}^2)$. We have then 
a full superfield analog of the gap equation involving the latter, including 
also the constant component and the $\theta^2$ component (with its 
conjugate). Potentially, one sees more interesting result options, like 
nontrivial solution from the $\theta^2$ part of the full superfield gap 
equation would give an alternative option of supersymmetry breaking.
\begin{figure}[!t]
\begin{center}

\includegraphics[scale=1]{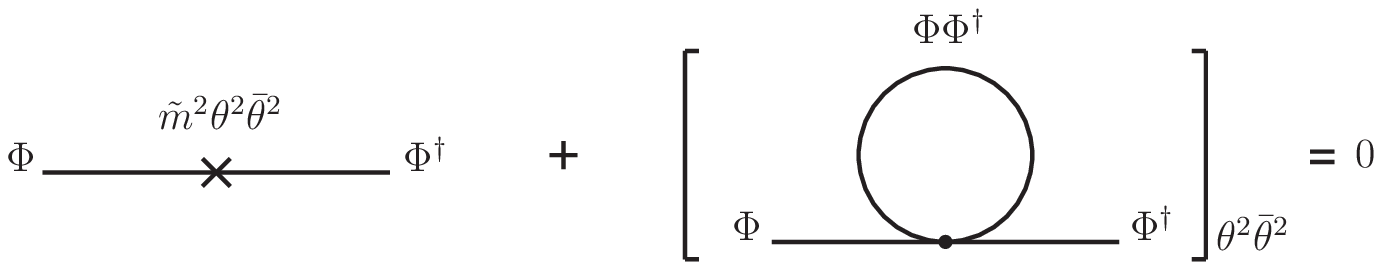}
\end{center}
\caption{\small  The soft mass gap equation in terms of the scalar component
$A$ of the superfield. }
\vspace*{.1in}
\hrule
\end{figure}

Before getting into our formulation, some comments on the symmetry
issues are in order. Apart from supersymmetry itself, the model
Lagrangian has, independent of the multiplet content of $\Phi_a$,
a $U(1)_{\!\ssc R}$ symmetry under which $\Phi_a$ has unit charge.
With vanishing $m_o$, it has a full $U(N)$ symmetry under which
the multiplet can be considered in the fundamental representation.
The $m_o =0$ case is really a main focus for us, though we do not
enforce it in the analysis to keep our result more general.  It is
important to note that a nonzero mass is not necessary for our key
result here, as presented below. The usual $1/N$ approximation
picture, however, can still be valid with the mass nonzero. $\Phi_a$ 
may then be considered as an $SO(N)$, instead of $SU(N)$, multiplet. 
In both cases, there is also $U(1)$ $\Phi$-number symmetry in the 
Lagrangian which is only violated by the mass term. In the naive case 
of really a single superfield, the gap equation analysis here would 
correspond to the quenched planar approximation of QED by Bardeen 
{\it et.al.} \cite{Bll,Bll1,Bll11}, which is commonly believed to give the
correct qualitative result in the kind of dynamical symmetry breaking 
studies. Some more discussion of the issue in a somewhat different 
setting is available in Ref.\cite{050}. 
\footnote{
For taking explicitly the single field case, there is then no difference in
the four-superfield interaction considered here comparing to the old SNJL
model. However, there are still a few key difference in our study
compared to that. First, the coupling has different signs. Second, we ask 
and answer a different question. We ask the question if a $\bar{\Phi} \Phi$ 
composite and condensate may form in the absence of supersymmetry 
breaking terms in the original Lagrangian. The SNJL work asked if a kind 
of $\Phi\Phi$ composite and condensate may form and got a sure no 
answer when there is no input supersymmetry breaking terms.
}

To keep notation simple, we will present our analysis here onwards with
the index suppressed, as if we are working on a single superfield. What we
have in mind is really a $N$-multiplet of the $SO(N)$ or $SU(N)$. To retrieve
result for a nontrivial $N$ is straightforward. The one-loop contribution
such as the one in $\Sigma_{\Phi\Phi^\dagger}(p;\theta^2, \bar{\theta}^2)$
or $\Sigma_{\tilde{m}}(p)$ will have to be multiplied by the factor $N$.

In the full superfield picture,
$\Sigma_{\Phi\Phi^\dagger}(p;\theta^2, \bar{\theta}^2)$ should expand as
\bea
\Sigma_{\Phi_{\!\ssc R}\Phi_{\!\ssc R}^\dagger}(p; \theta^2 \bar{\theta}^2)
= \Sigma_{r}(p)-  \Sigma_{\tilde{\eta}}(p)  \theta^2
 -  \bar{\Sigma}_{\tilde{\eta}^*}(p)  \bar{\theta}^2
  - \Sigma_{\tilde{m}}(p)   \theta^2 \bar{\theta}^2 \;.
\eea
The part $\Sigma_{\tilde{m}}$ in itself is like a proper self-energy
contribution to the scalar  but not the fermion component, hence soft
supersymmetry breaking. With $\Phi= A+\sqrt{2}\psi\theta +F \theta^2$,
it is a $AA^*$ vertex. The soft supersymmetry breaking mass $\tilde{m}^2$
as a superfield term is just the $\theta^2 \bar{\theta}^2$ component of
the kinetic term, to which
$\Sigma_{\Phi\Phi^\dagger}(p;\theta^2, \bar{\theta}^2)$
is the quantum correction to the latter. The part $\Sigma_{\tilde{\eta}}$
is somewhat less obvious. It is a proper vertex of $A F^*$, to be matched
to another mass parameter $\tilde{\eta}$; the $\tilde{\eta}A F^*$ term
gives another kind of soft supersymmetry breaking mass not usually
discussed in the literature.
\footnote{
Looking at the content of the superfield kinetic term, one sees that it is the
parameter for a $A F^*$ term. After elimination of the auxiliary component,
we have in general $|\tilde{\eta}|^2$ as an extra contribution to the scalar mass.
In the presence of superfield mass $m$, or rather ${\mathcal{M}}=m -\eta \theta^2$,
products of $\tilde{\eta} m$ and $\tilde{\eta} \eta$  may contribute to
other mass terms. We do not consider any nonzero $\eta$ here. Note that $m$, 
or ${\mathcal{M}}$ with zero $\eta$, is only an input parameter of the original 
supersymmetric Lagrangian. One can easily see that $\tilde{\eta} m$
contributes a $AA$ mass term after the elimination of the auxiliary
component $F$, giving mass-squared eigenvalues of
 $|m|^2 + |\tilde{\eta}|^2 +\tilde{m}^2 \pm 2 |\tilde{\eta}| |m|$
to the two (real) scalar states.
}
Lastly, the supersymmetric part $\Sigma_{r}$ gives only
a kinetic term, hence contributes to wavefunction renormalization.
It is then easy to appreciate that a consistent superfield treatment
of the standard NJL analysis should consider modifying the
superfield propagator to incorporate plausible nonperturbative 
parameter of the generic form given by
\beq
{\mathcal{Y}}= y-\tilde{\eta}_o\theta^2
-\tilde{\eta}^*_o \bar{\theta}^2-\tilde{m}^2_o\theta^2 \bar{\theta}^2 \;
\eeq
containing not only the $\tilde{m}^2$ part but also its supersymmetric
partners. We write here $\tilde{m}^2_o$ instead of $\tilde{m}^2$ as the
parameter is not the physical soft mass yet. The component $y$ contributes
a (supersymmetry) wavefunction renormalization factor which renormalizes
all mass parameters accordingly, as shown below explicitly. Notice that
generation of nontrivial $y$ breaks no symmetry while generation of
$\tilde{m}^2$ breaks only supersymmetry. Nonvanishing $\tilde{\eta}$
however breaks the $U(1)_{\!\ssc R}$ symmetry together with supersymmetry.

\begin{figure}[!t]
\begin{center}
\includegraphics[scale=1]{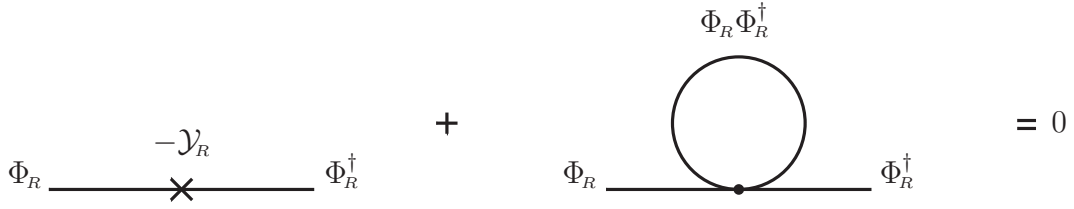}
\end{center}
\caption{\small  The renormalized superfield gap equation , with
${\mathcal{Y}_{\!\ssc R}} =\frac{y}{1+y}
-\tilde{\eta}\theta^2 -\tilde{\eta}^* \bar{\theta}^2
- \tilde{m}^2 \theta^2 \bar{\theta}^2$. }
\vspace*{.1in}
\hrule\label{srgap}
\end{figure}

To proceed with the derivation of the superfield gap equation,
we add and subtract the term ${\mathcal{Y}}   \bar{\Phi} \Phi$
and split the Lagrangian  as
${\mathcal{L}}={\mathcal{L}}_o +{\mathcal{L}}_{int}$ where
\bea
{\mathcal{L}}_o &=& \int  \bar{\Phi} \Phi (1+{\mathcal{Y}})
+\frac{m_o}{2}  \Phi^2 \delta^2(\bar{\theta})
+ \frac{m_o^*}{2}  \bar{\Phi}^2 \delta^2({\theta})
\eea
and
\bea
{\mathcal{L}}_{int} =   \int  - {\mathcal{Y}}  \bar{\Phi} \Phi
-\frac{g^2_o}{2}     \bar{\Phi} \Phi \bar{\Phi} \Phi  \;,
\eea
in which we have left the $d^4\theta$ implicit. To restore the canonical
kinetic term in the presence of a plausibly nonzero $y$, we introduce the
renormalized superfield $\Phi_{\!\ssc R}\equiv \sqrt{Z} \Phi=\sqrt{1+y} \Phi$
which gives
\bea
{\mathcal{L}}_o &=& \int  \bar{\Phi}_{\!\ssc R} \Phi_{\!\ssc R}
(1-\tilde{\eta}\theta^2
-\tilde{\eta}^* \bar{\theta}^2-\tilde{m}^2 \theta^2 \bar{\theta}^2)
+\frac{m}{2}  \Phi^2_{\!\ssc R} \delta^2(\bar{\theta})
+ \frac{m^*}{2}  \bar{\Phi}^2_{\!\ssc R} \delta^2({\theta})\;.
\eea
The mass parameters are of course renormalized ones, to be divided by
the wavefunction renormalization parameter $Z$; explicitly
$m=\frac{m_o}{1+y}$, for example. The quantum effective action is
\bea
\Gamma &=&   \bar{\Phi}_{\!\ssc R} \Phi_{\!\ssc R}
(1-\tilde{\eta}\theta^2
-\tilde{\eta}^* \bar{\theta}^2-\tilde{m}^2 \theta^2 \bar{\theta}^2)
+\frac{m}{2}  \Phi^2_{\!\ssc R} \delta^2(\bar{\theta})
+ \frac{m^*}{2}  \bar{\Phi}^2_{\!\ssc R} \delta^2({\theta})
\nonumber \\ && \hspace*{.2in}
-  {\mathcal{Y}}_{\!\ssc R}     \bar{\Phi}_{\!\ssc R} \Phi_{\!\ssc R}
-\frac{g^2}{2} \bar{\Phi}_{\!\ssc R} \Phi_{\!\ssc R} \bar{\Phi}_{\!\ssc R} \Phi_{\!\ssc R}
+\Sigma_{\Phi_{\!\ssc R}\Phi_{\!\ssc R}^\dagger} \bar{\Phi}_{\!\ssc R} \Phi_{\!\ssc R}
 + \cdots  \;,
\eea
where $g^2=\frac{g_o^2}{(1+y)^2}$ is the renormalized four-superfield
coupling and ${\mathcal{Y}}_{\!\ssc R}$ is similarly given by
\bea
{\mathcal{Y}}_{\!\ssc R} =
\frac{\mathcal{Y}}{Z} = \frac{y}{1+y}
-\tilde{\eta}\theta^2 -\tilde{\eta}^* \bar{\theta}^2
- \tilde{m}^2 \theta^2 \bar{\theta}^2 \;.
\eea
 The superfield gap equation under the NJL framework is then given by
\bea
- {\mathcal{Y}}_{\!\ssc R} + \left.
  \Sigma_{\Phi_{\!\ssc R}\Phi_{\!\ssc R}^\dagger}^{\tiny(loop)}(p; \theta^2 \bar{\theta}^2)
     \right|_{\mbox{\tiny on-shell}}   =0 \;;
\label{sgap}
\eea
in component form, we have
\bea
\frac{y}{1+y} &=& \left.\Sigma_{r}^{\tiny(loop)}(p)\right|_{\mbox{\tiny on-shell}}  \;,
\nonumber \\
\tilde{\eta} &=& \left.\Sigma_{\tilde{\eta}}^{\tiny(loop)}(p)\right|_{\mbox{\tiny on-shell}}  \;,
\nonumber \\
\tilde{m}^2 &=& \left.\Sigma_{\tilde{m}^2}^{\tiny(loop)}(p)\right|_{\mbox{\tiny on-shell}}  \;,
\label{cgap}
\eea
where in accordance of the standard NJL analysis one uses the one-loop
contribution to
$\Sigma_{\Phi_{\!\ssc R}\Phi_{\!\ssc R}^\dagger}(p; \theta^2 \bar{\theta}^2)$
from the four-superfield interaction. The diagrammatic illustration
of the renormalized superfield gap equation is given in
Fig.~\ref{srgap}. We can see that the naive expectation
of Eq.(\ref{gap}) works, so long as it understood to be applied to the
superfield and couplings with the wavefunction renormalization factor
properly incorporated. However, the wavefunction renormalization factor
itself can be retrieved from a gap equation. Note that results reported
in Ref.\cite{062} corresponds to assuming ${\tilde{\eta}}$ remains
zero from the beginning, which will be shown to be a consistent solution;
the gap equation figure therein is the $\theta^2 \bar{\theta}^2$ part
of the one here.

We perform a supergraph calculation for
$\Sigma_{\Phi_{\!\ssc R}\Phi_{\!\ssc R}^\dagger}(p;\theta, \bar{\theta})$ directly.
The relevant superfield propagator is given by
\bea
\langle T(\Phi(1)_{\!\ssc R} \Phi_{\!\ssc R}^\dagger(2)) \rangle &=&
\frac{-i}{p^2+|m|^2} \delta^4_{\ssc 12}
- i\frac{  \tilde\eta (Q-2|m|^2)}{Q^2- 4\,|m|^2|\tilde\eta|^2 } \, {\theta_{\!\ssc 1}}^2 \delta^4_{\ssc 12}
- i\frac{  \tilde\eta^* (Q-2|m|^2)}{Q^2- 4\,|m|^2|\tilde\eta|^2 } \, \bar{\theta_{\!\ssc 1}}^2 \delta^4_{\ssc 12}
\nonumber \\ &&
+  i\frac{(\tilde{m}^2 + |\tilde\eta|^2)Q - 4 |m|^2|\tilde\eta|^2}{(p^2+|m|^2|)(Q^2- 4\,|m|^2|\tilde\eta|^2) }
  \left[ \frac{D_{\!\ssc 1}^2 \theta_{\!\ssc 1}^2 \bar{\theta_{\!\ssc 1}}^2 \overline{D}_{\!\ssc 1}^2}
    {16} \right] \delta^4_{\ssc 12}
\nonumber \\ &&
+ i\frac{(-p^2|\tilde\eta|^2+ \tilde{m}^2|m|^2)Q + 4p^2 |m|^2|\tilde\eta|^2}{(p^2+|m|^2|)(Q^2- 4\,|m|^2|\tilde\eta|^2) } \,
\theta_{\!\ssc 1}^2 \bar{\theta_{\!\ssc 1}}^2 \delta^4_{\ssc 12} \;,
\label{sprop}
\eea
where  $Q=p^2+|m|^2+|\tilde\eta|^2+ \tilde{m}^2$ and
$\delta^4_{\ssc 12}=\delta^4(\theta_{\!\ssc 1}-\theta_{\!\ssc 2})$.
The necessary evaluation of
$\left. \Sigma_{\Phi_{\!\ssc R}\Phi_{\!\ssc R}^\dagger}^{\tiny(loop)}
(p; \theta^2 \bar{\theta}^2) \right|_{\mbox{\tiny on-shell}}$
is much to similar previous cases \cite{042}. The result is given by
\bea
\left. \Sigma_{\Phi_{\!\ssc R}\Phi_{\!\ssc R}^\dagger}^{\tiny(loop)}
(p; \theta^2 \bar{\theta}^2) \right|_{\mbox{\tiny on-shell}}
&=& - g^2 \int^{\!\ssc E} \left[
\frac{1}{k^2+|m|^2}
+ \frac{  \tilde\eta (Q_k-2|m|^2)}{Q_k^2- 4\,|m|^2|\tilde\eta|^2 } \, {\theta}^2
+ \frac{  \tilde\eta^* (Q_k-2|m|^2)}{Q_k^2- 4\,|m|^2|\tilde\eta|^2 } \, \bar{\theta}^2
\right. \nonumber \\  &&
- \frac{(\tilde{m}^2 + |\tilde\eta|^2)Q_k - 4 |m|^2|\tilde\eta|^2}{(k^2+|m|^2|)(Q_k^2- 4\,|m|^2|\tilde\eta|^2) }
\left( 1- k^2 \theta^2 \bar{\theta}^2+4 k_a \sigma^a_{\alpha\dot{\alpha}}
\theta^{\alpha} \bar{\theta}^{\dot{\alpha}} \right)
\nonumber \\  && \left.
- \frac{(-k^2|\tilde\eta|^2+ \tilde{m}^2|m|^2)Q_k + 4k^2 |m|^2|\tilde\eta|^2}{(k^2+|m|^2|)(Q_k^2- 4\,|m|^2|\tilde\eta|^2) } \,
\theta^2 \bar{\theta}^2
\right] \;,
\eea
where the $\int^{\!\ssc E}$  denotes integration over Euclidean
four-momentum $k$ with the measure $\frac{d^4k}{(2\pi)^4}$
and $Q_k=k^2+|m|^2+|\tilde\eta|^2+ \tilde{m}^2$. Each of the five 
terms in the above expression comes exactly from the  
corresponding term in the superfield propagator. The
$4 k_a \sigma^a_{\alpha\dot{\alpha}}\theta^{\alpha} \bar{\theta}^{\dot{\alpha}}$
term vanishes upon integration. The others can be pull together 
to give the component gap equations as
\bea
\frac{y}{1+y}&=&
\left.\Sigma_{r}^{\tiny(loop)}(p)\right|_{\mbox{\tiny on-shell}} =
-g^2  \int^{\!\ssc E}  \frac{( k^2+ |m|^2 + \tilde{m}^2+ |\tilde\eta|^2 )}
   { ( k^2+ |m|^2 + \tilde{m}^2+ |\tilde\eta|^2)^2 - 4|m|^2 |\tilde{\eta}|^2 } \;,
\nonumber \\
\tilde{\eta} &=&
\left.\Sigma_{\tilde{\eta}}^{\tiny(loop)}(p)\right|_{\mbox{\tiny on-shell}}  =
 g^2  \tilde{\eta} \int^{\!\ssc E}  \frac{( k^2- |m|^2 + \tilde{m}^2+ |\tilde\eta|^2 )}
   { ( k^2+ |m|^2 + \tilde{m}^2+ |\tilde\eta|^2)^2 - 4|m|^2 |\tilde{\eta}|^2 } \;,
\nonumber \\
\tilde{m}^2 &=&
\left.\Sigma_{\tilde{m}^2}^{\tiny(loop)}(p)\right|_{\mbox{\tiny on-shell}}  =
g^2  \int^{\!\ssc E}
\frac{1}{(k^2+|m|^2)}
\frac{1}{( k^2+ |m|^2 + \tilde{m}^2+ |\tilde\eta|^2)^2 - 4|m|^2 |\tilde{\eta}|^2}
\nonumber\\&&
\cdot \left\{ \left[\tilde{m}^2(k^2-|m|^2)+2 k^2|\tilde\eta|^2 \right]
(k^2+|m|^2+\tilde{m}^2+|\tilde\eta|^2) - 8 k^2|m|^2
|\tilde\eta|^2 \right\} \,.
\label{soln}
\eea
Nontrivial solutions of the three coupled equations with nonvanishing
$\tilde{\eta}$ and/or $\tilde{m}^2$ give supersymmetry breaking 
solutions. We postpone the analysis of the nontrivial solution till after 
the discussion of the effective theory picture in the next section. Note 
that nontrivial $y$ value gives wavefunction renormalization to $\Phi$
which does not change the qualitative answer to if supersymmetry
breaking solution with the soft mass generation exists. Our analysis
will explicitly demonstrate that.

\section{The Effective Theory Picture}
Following the general effective theory picture of the NJL-type models,
we modify the model Lagrangian by adding to it
\beq
\mathcal{L}_s =  \int d^{4} \,\theta \frac{1}{2}(\mu U + g_o \bar{\Phi} \Phi )^2 \;,
\eeq
where $U$ is an `auxiliary' real superfield and mass parameter $\mu$ taken as
real and positive (for $g_o^2>0$). The equation of motion for $U$, from the full
Lagrangian $\mathcal{L} +\mathcal{L}_s$ gives
\beq
U = - \frac{g_o}{\mu} \bar{\Phi} \Phi \;,
\eeq
showing it as a superfield composite of $\bar{\Phi}$ and $\Phi$.
The condition says the model with $\mathcal{L} +\mathcal{L}_s$ is
equivalent to that of $\mathcal{L}$ alone. Expanding the term in
$\mathcal{L}_s$, we have a cancellation of the dimension six interaction
in the full Lagrangian, giving it as
\beq
\mathcal{L}_{ef\!f}\equiv \mathcal{L} +\mathcal{L}_s
 = \int d^{4}\theta \left[ \bar{\Phi}  \Phi
 + \frac{\mu^2}{2} U^{2} + \mu g_o U \bar{\Phi}\Phi
+\frac{m_o}{2} \Phi^2 \delta(\bar{\theta})
+ \frac{m_o^*}{2}  \bar{\Phi}^2 \delta(\theta) \right] \;.
\label{Le}
\eeq
Obviously, if $\left. U \right|_D$ develops a vacuum expectation value
(VEV), supersymmetry is broken spontaneously and the superfield $\Phi$
gains a soft supersymmetry breaking mass of
$\tilde{m}^2_o=-\mu g_o \lla\left. U \right|_D\rra$.
The above looks very much like the standard features of NJL-type model.
Notice that while $U$ does contain a vector component, its couplings
differ from that of the usually studied `vector superfield' which is
a gauge field supermultiplet. That is in addition to having $\mu$ as like
a supersymmetric mass for $U$, which can be compatible only with a
broken gauge symmetry. As such, model with superfield $U$ is not
usually discussed. The superfield can be seen as two parts, as
illustrated by the following component expansion,
\begin{eqnarray}
U(x,\theta,\bar\theta ) &=&  \frac{C(x)}{\mu}
+\sqrt{2}\theta \frac{\chi(x)}{\mu} + \sqrt{2} \bar\theta \frac{\bar\chi(x)}{\mu}
+\theta \theta \frac{N(x)}{\mu} +\bar\theta \bar\theta \frac{{N^*}(x)}{\mu}
\nonumber \\  && \;\;\;
+\sqrt{2}\theta \sigma^{\mu}\bar \theta v_{\mu}(x)
+\sqrt{2}\theta \theta \bar\theta  \bar\lambda(x)
+\sqrt{2}\bar\theta \bar\theta \theta   \lambda(x)
+\theta \theta \bar\theta \bar\theta D(x)\; ,
\end{eqnarray}
where the components $C$, $\chi$, and $N$ is the first part
which has the content of like a chiral superfield with however $C$
being real. The $\mu$ factor is put to set the mass dimensions
right. The rest is like the content of a superfield for the usual
gauge field supermultiplet, with $D$ and $v_\mu$ real. The effective
Lagrangian in component form is given by
\bea
{\cal L}_{ef\!f}
&=&
(1+g_o C)\left[
          A^*\Box A
         +i(\partial_\mu\bar{\psi}) \bar{\sigma}^\mu\psi
         + F^* F   \right]
+\frac{m_o}{2}   \left( 2 A F  -\psi\psi  \right)
+\frac{m^*_o}{2}  \left(
   2 A^* F^*  -\bar{\psi}\bar{\psi} \right)
\nonumber\\  &&
+ {\mu}CD  -{\mu}\chi\lambda -{\mu}\bar{\chi}\bar\lambda
+NN^* -\frac{\mu^2}{2}v^{\nu}v_{\nu}
-\mu g_o\psi \lambda A^* - \mu g_o\bar{\psi}\bar{\lambda} A  + \mu g_o D   A^* A
\nonumber\\  &&
-i\frac{g_o}{2}\bar{\psi}\bar{\sigma}^\mu\chi \partial_\mu A
+i\frac{g_o}{2}(\partial_\mu\bar{\psi})\bar{\sigma}^\mu\chi A
-g_o\chi\psi F^* +g_o N A F^*
\nonumber\\  &&
+i\frac{g_o}{2}\bar{\chi}   \bar{\sigma}^\mu\psi  \partial_\mu A^*
-i\frac{g_o}{2} A^*   \bar{\chi}\bar{\sigma}^\mu \partial_\mu\psi
-g_o\bar{\chi}\bar{\psi}F +g_o N^* A^* F
\nonumber\\  &&
- \frac{\mu g_o}{\sqrt{2}}\eta^{\mu\nu}v_\mu iA^*\partial_\nu A
  + \frac{\mu g_o}{\sqrt{2}}\eta^{\mu\nu}v_\mu i(\partial_\nu A^*)A
  -\frac{\mu g_o}{\sqrt{2}}\eta^{\mu\nu}v_\mu \bar{\psi}\bar{\sigma}_\nu\psi
\;. \label{Lec}
\eea
Notice that like $F,$ $N$ and $D$ have mass dimension two.

Under the  $U(1)_{\!\ssc R}$ symmetry, $A$ and $F$ have charge $+1$ and $-1$.
The superfield $U$ is uncharged. However, components $N$, $\chi$ and $\lambda$
carry nontrivial $U(1)_{\!\ssc R}$ charges -2, -1 and +1, respectively.
For the $m_o=0$ case, there is an extra $U(1)$ $\Phi$-number symmetry
with common charge for all components. All components of $U$ is not
charged under the latter.

In accordance with the `quark-loop' approximation in the (standard) NJL
gap equation analysis and our particular supergraph calculation scheme
above in particular, we consider plausible nontrivial vacuum solution with 
nonzero vacuum expectation values (VEVs) for the composite scalars $C$, 
$D$ and $N$. While $N$ is complex, we can safely taken $n\equiv\lla N \rra$
to be real here. At least we can exploit the $U(1)_{\!\ssc R}$ symmetry
to absorb any phase at the expense of having a complex $m_o$ the phase
of which does not show up in the calculation. First note that scalar $C$ couples 
to kinetic terms of components of $\Phi$; $c\equiv\lla C \rra$ hence 
contributes to a supersymmetric wavefunction renormalization of the latter.
It is the supersymmetric part of
 $\Sigma^{({\mbox \tiny loop})}_{\Phi\Phi^\dagger}(p;\theta^2, \bar{\theta}^2)$
an unavoidable part of the one-loop supergraph in our gap equation
calculation in the previous section. Again, we should go to the renormalized
superfield $\Phi_{\!\ssc R}= \sqrt{(1+g_o c)} \Phi$ in the following
calculations, with renormalized mass $m$ and coupling $g$.
With $n\equiv\lla N \rra$  and $d\equiv\lla D \rra$, we have $-g n$
and ${-\mu gd}$ corresponding to the supersymmetry breaking masses
$\tilde{\eta}$ and $\tilde{m}^2$ of $\Phi_{\!\ssc R}$. In the former case,
it gives a $A_{\!\ssc R}F_{\!\ssc R}^*$ component term. Note that $\lla N \rra$
is the only VEV that breaks the $U(1)_{\!\ssc R}$ symmetry, as $C$ and $D$ carry
no charges, though both $\lla N \rra$ and $\lla D \rra$ break supersymmetry.

\begin{figure}[!t]
\begin{center}
\includegraphics[scale=1]{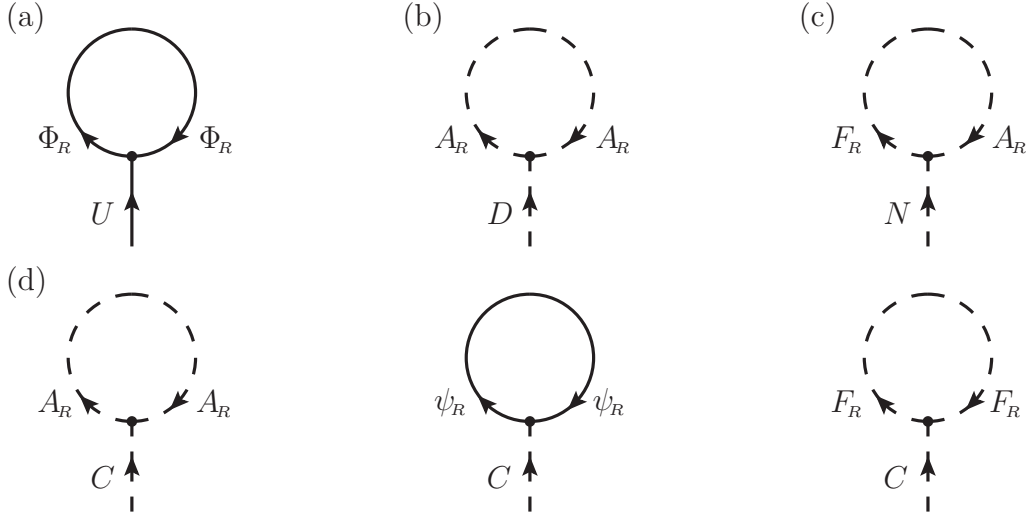}
\end{center}
\caption{\small  The tadpole diagrams: a) the superfield diagram;
b) D-tadpole; c) N-tadpole; d) C-tadpoles. %
}
\vspace*{.1in}
\hrule
\label{tp}
\end{figure}

With propagators for the components of the renormalized `quark' superfield
$\Phi_{\!\ssc R}$ as given  in the appendix, one can easily obtain the
minimum condition for the effective potential following the Weinberg
tadpole method \cite{W,M}. Firstly, for $C$-tadpoles, we have a
$\Phi_{\!\ssc R}$ loop or in component form one from each of $A_{\!\ssc R}$,
$\psi_{\!\ssc R}$, and $F_{\!\ssc R}$. Hence, we have up to one loop level
\bea
\Gamma^{(1)}_{C} = \Gamma^{(1)\mbox{\tiny tree}}_{C}
  + \Gamma^{(1)}_{C_{A}} +\Gamma^{(1)}_{C_\psi}
  + \Gamma^{(1)}_{C_F}
 = {\mu}  d  - g I_{\!\ssc C}  \;,
\eea
where
\bea
I_{\!\ssc C} &=&  I_{\!\ssc C\!A} - 2 I_{\!\ssc C\psi}
   + I_{\!\ssc C\!F} \;;
\nonumber \\
I_{\!\ssc C\!A} &=&  \int^{\!\ssc E} 
   \frac{k^2 (k^2+ |m|^2 + g^2 |n|^2 -\mu g d)}
   { ( k^2+ |m|^2 + g^2 |n|^2 -\mu g d)^2 - 4 g^2 |n|^2 |m|^2  }\;,
\nonumber \\
I_{\!\ssc C\psi} &=& \int^{\!\ssc E} 
   \frac {k^2}{  k^2+|m|^2  }    \;,
\nonumber \\
I_{\!\ssc C\!F} &=&  \int^{\!\ssc E} 
    \frac{(k^2  -\mu g d) ( k^2+ |m|^2 + g^2 |n|^2 -\mu g d)}
   { ( k^2+ |m|^2 + g^2 |n|^2 -\mu g d)^2 - 4 g^2 |n|^2 |m|^2  }    \;.
\eea
Next, the $N^*$-tadpole is given by
\bea
\Gamma^{(1)}_{N^*}&=& n- g I_{\!\ssc N}   \;,
\eea
where
\bea
I_{\!\ssc N} &=&   \int^{\!\ssc E} 
\frac{ gn(k^2- |m|^2 + g^2 |n|^2 -\mu g d)}
   { ( k^2+ |m|^2 + g^2 |n|^2 -\mu g d)^2 - 4 g^2 |n|^2 |m|^2  }\;.
\eea
The $D$-tadpole is given by
\bea
\Gamma^{(1)}_{D}&=&{\mu} c  +  {\mu g}\,I_{\!\ssc D}
\eea
where
\bea
I_{\!\ssc D} &=& \int^{\!\ssc E} 
   \frac{k^2+ |m|^2 + g^2 |n|^2 -\mu g d}
   { ( k^2+ |m|^2 + g^2 |n|^2 -\mu g d)^2 - 4 g^2 |n|^2 |m|^2  }\;.
\eea
The tadpole diagrams are illustrated in Fig.~\ref{tp}. We look for vacuum solution 
with $-\Gamma^{(1)}_{a}\equiv \partial V(c,n,d)_{\mbox{\tiny 1-loop}}/\partial a=0$
for $a=c, n, d$. Firstly,  note that the vanishing of $N^*$-tadpole is equivalent to
\bea
n(1-g^2 I_{\!\ssc N'})=0 \;,
\eea
with $I_{\!\ssc N'}$ given by $I_{\!\ssc N}= gn I_{\!\ssc N'}$;
vanishing $D$-tadpole gives
\bea
c = -g  I_{\!\ssc D} \;;
\eea
the vanishing $C$-tadpole condition is
\bea
\mu d =g I_{\!\ssc C} \;.
\eea
To get the physics picture clear, one can identify the soft masses generated 
for the superfield $\Phi$ by $\tilde{\eta}=- g n$ and $\tilde{m}^2= -\mu g d$. 
We will explore nontrivial solutions for the soft masses below.

It is interesting to see that the effective potential analysis for (the
components of) the composite superfield $U$ can be shown directly to be
equivalent to the superfield gap equation, which we illustrated explicitly
in Ref.\cite{062} and duplicated here. In terms of the superfield,
the potential minimum condition is given by
\bea &&
\mu^2 \lla U \rra + U_{tadpole} =0
 \qquad    \Longrightarrow \qquad
\mu g \lla U \rra =  - g^2 I^{\tiny(loop)}_{\Phi_{\!\ssc R}\Phi_{\!\ssc R}^\dagger}
\eea
where $I^{\tiny(loop)}_{\Phi_{\!\ssc R}\Phi_{\!\ssc R}^\dagger}$ is the 
momentum integral of the ${\Phi_{\!\ssc R}\Phi_{\!\ssc R}^\dagger}$ 
propagator loop ({\em cf.} the first diagram in Fig~\ref{tp}). Note that from 
the original Lagrangian with two-superfield composite assumed, we can obtained
$-g^2 \lla \left( \Phi_{\!\ssc R}\Phi_{\!\ssc R}^\dagger \right) \rra
 ={{\mathcal{Y}}_{\!\ssc R}}$,
which is equivalent to $\mu g \lla U \rra = {{\mathcal{Y}}_{\!\ssc R}}
= \left. \Sigma_{\Phi_{\!\ssc R}\Phi_{\!\ssc R}^\dagger}^{\tiny(loop)}(p; \theta^2 \bar{\theta}^2)  \right|_{\mbox{\tiny on-shell}}
 =  - g^2 I^{\tiny(loop)}_{\Phi_{\!\ssc R}\Phi_{\!\ssc R}^\dagger}$.
The same loop integral is of course involved in both the gap equation
picture and the effective potential analysis. The results here are in direct
matching with the corresponding discussion for the NJL case presented in
Ref.\cite{BE}, though for a superfield theory instead. The component
field effective potential analysis here above hence really serves as
a double-check of the superfield gap equation analysis of the previous
section. In terms of component fields, we need the soft mass 
identifications above as well as $y=g_o c$, or $\frac{y}{1+y}=g c$.

\section{Supersymmetry Breaking Solutions}
Let us pull together the gap equation in terms of  $y=(1-gc)^{-1}$,
$\tilde{\eta}(=- g n)$ and $\tilde{m}^2(= -\mu g d)$. We have
\[
\tilde{\eta} (1-g^2 I_{\!\ssc N}') =0
\quad \mbox{and}
\quad
\tilde{m}^2 = -g^2 I_{\!\ssc C} \;,
\]
as a set of coupled equations to be solved simultaneously as
the integrals are complicated expressions involving the two
soft mass parameters. The third equation of
\[
y=(1+g^2 I_{\!\ssc D})^{-1}
\]
independently gives the $y$ value for any solution of $\tilde{\eta}$
and $\tilde{m}$. One can easily check that the equations are indeed
identical to the set of Eq.(\ref{soln}) derived from the original Lagrangian 
through the supergraph evaluation. Note that the $y$ parameter does not 
correspond to any physical quantity and hence may be considered of little 
interest. The case of zero soft masses is consistent, as $I_{\!\ssc C}$  vanishes 
in the supersymmetric limit. The point of interest is if solutions of nontrivial 
supersymmetry breaking masses $\tilde{\eta}$ and $\tilde{m}$ exist.

\begin{figure}[!t]
\begin{center}
\includegraphics[height=8.0cm,width=10.0cm]{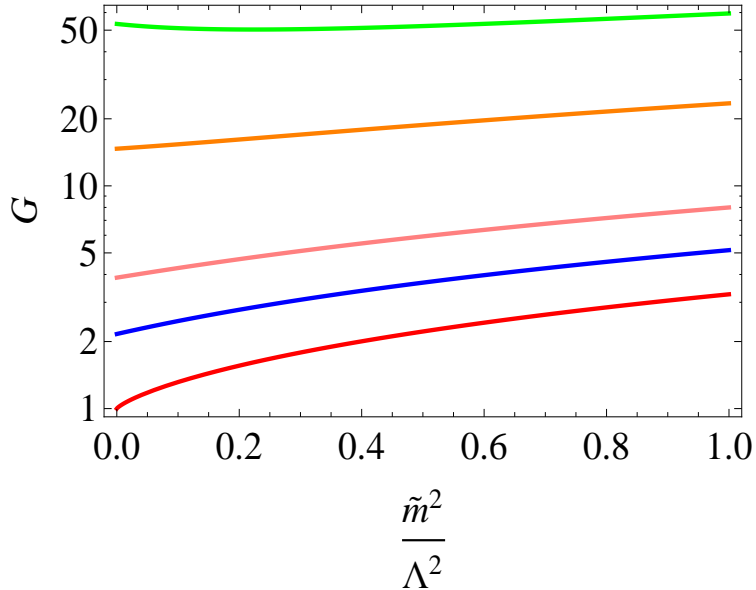}
\end{center}
\caption{\small Numerical plot  of nontrivial solutions to the soft mass
gap equation with $|\tilde\eta|=0$. Coupling parameter 
$G=\frac{N g^2  \Lambda^2}{16\pi^2}$ is plotted against the normalized 
soft mass parameter $s\;\left(=\frac{\tilde{m}^2}{\Lambda^2}\right)$
for  $t\;\left(=\frac{{|m|}^2}{\Lambda^2}\right)$ values of $0$ (red),
$0.1$ (blue), $0.2$ (pink), $0.4$ (orange), $0.5$ (green), from the lowest to
the highest curves, respectively. Here $N$ is the `color' factor for the case 
of the basic chiral superfield $\Phi$ being an $SO(N)$ or $SU(N)$ multiplet
not shown explicitly in the calculation, and $\Lambda$ is the model cutoff 
scale. Notice that the critical coupling increases from $G=1$ for nonzero
values of the input supersymmetric mass $m$. (Figure duplicated from
Ref.\cite{062}.)}
\vspace*{.1in}
\hrule
\label{plot}
\end{figure}

The first soft mass gap equation gives $g^2 I_{\!\ssc N}'=1$ for nontrivial
$\tilde{\eta}$, for the case of which we have
\bea
 I_{\!\ssc N}' = \frac{1}{2} \left[ \left(1-\frac{|m|}{|\tilde{\eta}|}\right)  I_{\!\ssc F}(m_{\!\ssc A_-}^2)
    +  \left(1+\frac{|m|}{|\tilde{\eta}|}\right)  I_{\!\ssc F}(m_{\!\ssc A_+}^2) \right] \;,
\label{gIn'}\eea
where $I_{\!\ssc F}(S) \left[ \equiv\int^{\ssc E} \frac{1}{k^2+S} \right]$
has been used to denote integral of the Feynman propagator
for field of mass square $S$ and we have the scalar mass eigenvalues
\footnote{
In connection to the scalar masses, it is interesting to note that nontrivial
$\tilde{\eta}$ also gives spontaneous CP violation. Though we keep $m$ as a 
complex parameter in our analysis, its complex phase in the original Lagrangian 
is not physical and can be taken away. The original Lagrangian hence conserves 
CP. Or as seen here, presence of nonzero $m\tilde{\eta}$ product splits the masses
of the scalar and pseudoscalar part of $A$ and produces mass mixing between them,
giving the mass eigenvalues. The surprising part is one only needs nonzero
$m\tilde{\eta}$ to have it, even real value would do.}
\bea
m_{\!\ssc A_\mp}^2 &=& \tilde{m}^2 +  (|m| \mp |\tilde{\eta}|)^2 \;.
\label{mA}
\eea
Similarly, we have
\bea
 I_{\!\ssc C} = -\left(m_{\!\ssc A_-}^2 -\frac{\tilde{m}^2}{2}\right)
 I_{\!\ssc F}(m_{\!\ssc A_-}^2) - \left(m_{\!\ssc A_+}^2 -\frac{\tilde{m}^2}{2}\right) I_{\!\ssc F}(m_{\!\ssc A_+}^2)
+ 2 |m|^2 I_{\!\ssc F}(|m|^2) \;.
\label{gIc}\eea
If we take $m=0$,  we would have
\[
I_{\!\ssc N}'  \longrightarrow  I_{\!\ssc F}(|\tilde{\eta}|^2+\tilde{m}^2)
\]
and
\[
 I_{\!\ssc C} \longrightarrow -\tilde{m}^2 I_{\!\ssc F}(|\tilde{\eta}|^2+\tilde{m}^2)
- 2|\tilde{\eta}|^2  I_{\!\ssc F}(|\tilde{\eta}|^2+\tilde{m}^2) \;.
\]
The second soft mass gap equation becomes
\bea
g^2  I_{\!\ssc F}(|\tilde{\eta}|^2+\tilde{m}^2)
 \left(1+  2 \frac{|\tilde{\eta}|^2}{\tilde{m}^2} \right) =1
\eea
which is not compatible with the first one ($g^2 I_{\!\ssc N}'=1)$ unless
$\tilde{\eta}=0$. It remains to see if there exists $\tilde{\eta} \ne 0$
solution for some nonzero values of $m$. After some algebra, one can 
rewrite the solution equations in the form
\bea
g^2 I_{\!\ssc F}(m_{\!\ssc A_\mp}^2) &=&
 \frac{|m|\tilde m^2 \mp 2|\tilde\eta|(|m|\pm|\tilde\eta|)^2}
{|m|(2 |m|^2 -2|\tilde\eta|^2+\tilde m^2)}
+\frac{2|m|(|m|\pm|\tilde\eta|)}{2 |m|^2 -2|\tilde\eta|^2+\tilde m^2 }
g^2 I_{\!\ssc F}(|m|^2) \;.
\eea
The two equations have the same form with only the $|\tilde\eta|$
variable come in different signs.  And both reduces to the same equation
for the $I_{\!\ssc F}(m_{\!\ssc A})$ at the $|\tilde\eta|=0$ limit, which
is the gap equation for the limiting case \cite{062}. Evaluating the integrals 
with model cutoff $\Lambda$, with all variables and parameters casted in 
terms of dimensionless counterparts normalized to $\Lambda$ given by
$G=\frac{g^2  \Lambda^2}{16\pi^2}$, $s=\frac{\tilde{m}^2}{\Lambda^2}$,
and $t=\frac{|m|^2}{\Lambda^2}$, the two equations are equivalent to
\bea
\frac{1}{G(s,t,z)} &=&
 \frac{s+2tz(1-z)}{s+2tz(1-z)^2}
+ \frac{2t(1-z)}{s+2tz(1-z)^2} \, t\ln\left[1+\frac{1}{t}\right]
\nonumber\\ &&
-\frac{s+2t(1-z^2)}{s+2tz(1-z)^2} \, [s+t(1+z)^2]\ln\left[1+\frac{1}{s+t(1+z)^2} \right] \;,
\label{Gpm}
\eea
for  $z=\mp\frac{|\tilde\eta|}{|m|}$ respectively. We need simultaneous 
solutions for $s$ and $z$ for reasonable values of model parameters $G$ 
and $t$. The two equations for positive and negative (but equal) values of 
$z$ of course collapse to one at $z=0$, which is the vanishing
$|\tilde\eta|$ solutions which we presented in Ref.\cite{062}. 
We duplicate the illustrating plots here in Fig.~\ref{plot}.

Actually, in the $\tilde{\eta} =0$ $(z=0)$ case, all the above
integrals simplify analytically. In particular, we have
\[
 I_{\!\ssc D} \longrightarrow  I_{\!\ssc F}(|m|^2+\tilde{m}^2)
\]
and
\[
 I_{\!\ssc C} \longrightarrow  -(\tilde{m}^2+2|m|^2)  I_{\!\ssc F}(|m|^2+\tilde{m}^2)
+2|m|^2  I_{\!\ssc F}(|m|^2)
\]
($I_{\!\ssc N}'$ is irrelevant). The masses in the Feynman propagators
correspond to the scalar and fermion masses.  It is interesting to note
that for $m=0$ $(t=0)$, we have the simple result
$g^2  I_{\!\ssc F}(\tilde{m}^2) =1$
\footnote{It is interesting to note that in the case we have the gap 
equation for the renormalization factor, which is equivalent to the 
vanishing $D$-tadpole condition $c=-gI_{\!\ssc D}$,  giving 
$c=-g I_{\!\ssc F}(\tilde{m}^2)$ hence $g_o \,c= -.5$. The wavefunction 
renormalization factor is $Z= 1+ g_o \,c=.5$, of order one but tangible.
That is a clear indication of the nonperturbative nature of the
results and that there is nothing improper in the analysis.
},
which is the same as the basic NJL model one except with the soft mass 
$\tilde{m}^2$ replacing the (Dirac) fermionic mass (see for example 
Ref.\cite{BE}) if we take $\frac{g^2}{2}$ as the four-fermion coupling in 
the model. For more details, we see that solutions for nontrivial 
$\tilde{m}^2$  for the case is given by the reduced form of Eq.(\ref{Gpm}) as
 $\frac{1}{G} = 1 - s \ln\left[1+\frac{1}{s}\right]$ obviously giving solution 
for $0<s<1$ for the strong enough coupling $G>1$. It can be seen from the 
numerical plot that the value of the $\tilde{m}^2$ solution rises fast with 
increasing $G$. However, nonzero $t$ has a strong limiting effect. It 
increases the critical coupling needed for a nontrivial solution to $s$ 
very substantially. In fact, taking the limit $s\to 0$, the equation becomes
$\frac{1}{G_c} = 1 + \frac{2t}{t+1}- 3t \ln\left(1+\frac{1}{t}\right)$,
which gives the critical coupling ${G_c}$  as a function of $t$ . It can be 
seen then as $t$ increases from zero,  $\frac{1}{G_c}$ decreases and 
reaches vanishing value ({\em i.e.} $G_c\to \infty$) at a critical $t$ value 
of about $0.55$, beyond which no coupling $G$ will be strong enough to 
break the supersymmetry and generate the soft mass. 
This part of the results has been reported in Ref.\cite{062}.

Looking for solution with nontrivial $\tilde{\eta}$ is more tricky and 
requires a very careful analysis scanning the numerical results. Again, we 
check plots of the effective coupling $G$ as given in Eq.(\ref{Gpm}) versus 
$s$ simultaneously for positive and negative values of $z$ of fixed 
magnitude, at a fixed input $t$ value. Numerically, where the two curves 
(dubbed $G_+$ and $G_-$, respectively) intersect within the window of 
interest gives a solution. One only have then to numerically scan the plots 
of the $G_+$ and $G_-$ curves to see all the solutions. The window of 
interest is restricted to positive $G$ value and $0<s\leq 1$ plus the extra 
constraint of both of the mass eigenvalues of the scalar states in $\Phi$ to 
be within the cutoff $\Lambda$ [{\em cf.} Eq.(\ref{mA})]. This is  the 
generalization of  $s\leq 1$ to the nontrivial $\tilde{\eta}$, $|z|\ne 0$ 
case. The constraint is given by
\bea \label{sb}
s+t(1+|z|)^2 \leq 1 \;.
\eea
It is strong. For any $t$ value, it first restricts $|z|$ of interest to 
$\leq \frac{1}{\sqrt{t}} -1$.  Close to the upper limit means $s$ admissible has 
to be very small. So, the constraint may cut out quite a range, if not all, of the 
$s$ value of interest. We find that solution exists in general, though some 
of the features of the solution locations are not somewhat peculiar and not 
easy to understand.

\begin{figure}[!t]
\begin{center}
\includegraphics[height=6.0cm,width=10.0cm]{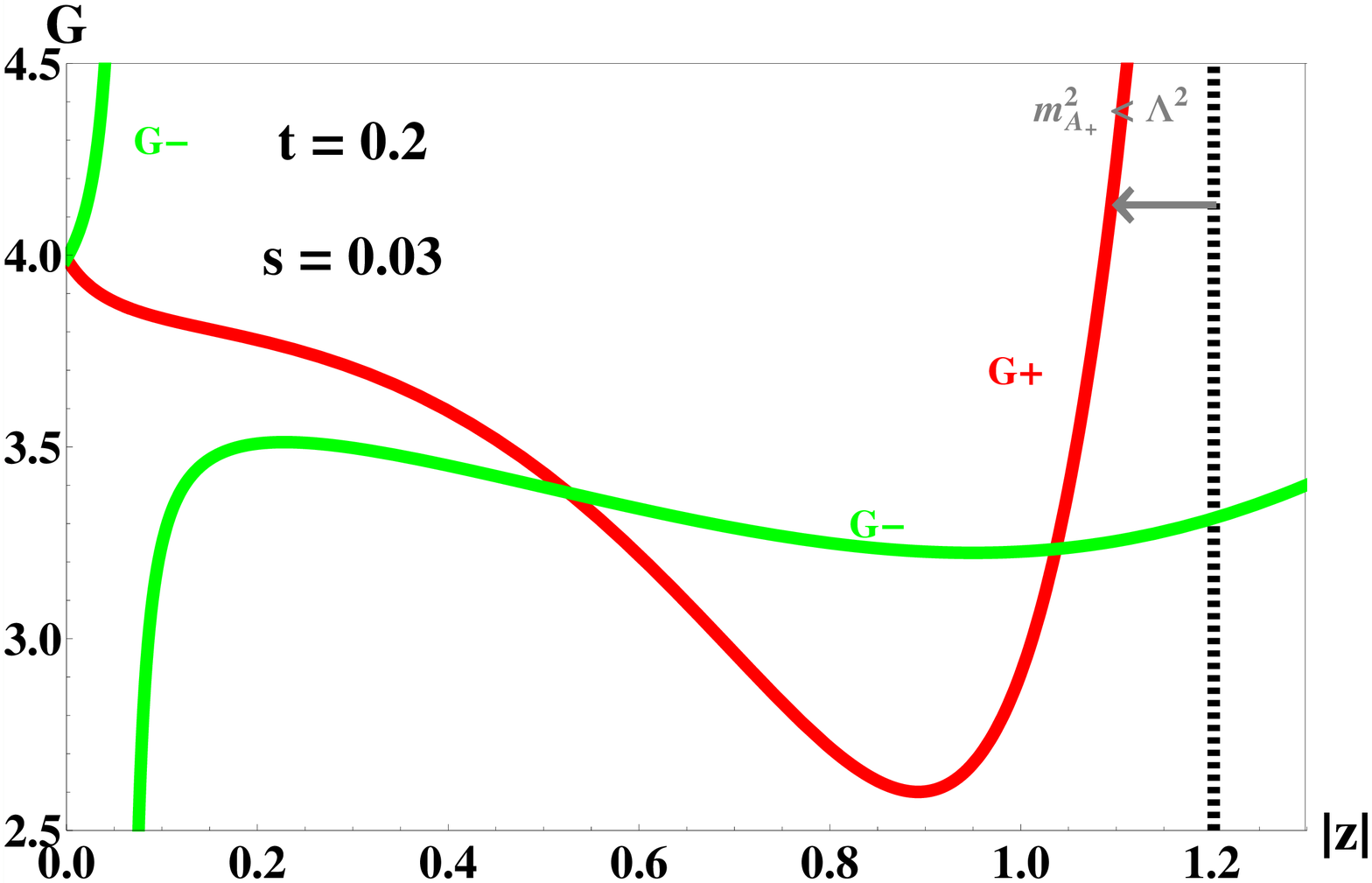}
\end{center}
\caption{\small  An illustrative  of intersecting point solutions,
with $G$  versus $|z|$.
 }
\vspace*{.1in}
\hrule
\label{Gz}
\end{figure}
\begin{figure}[!t]
\begin{center}
\includegraphics[height=8.0cm,width=8.0cm]{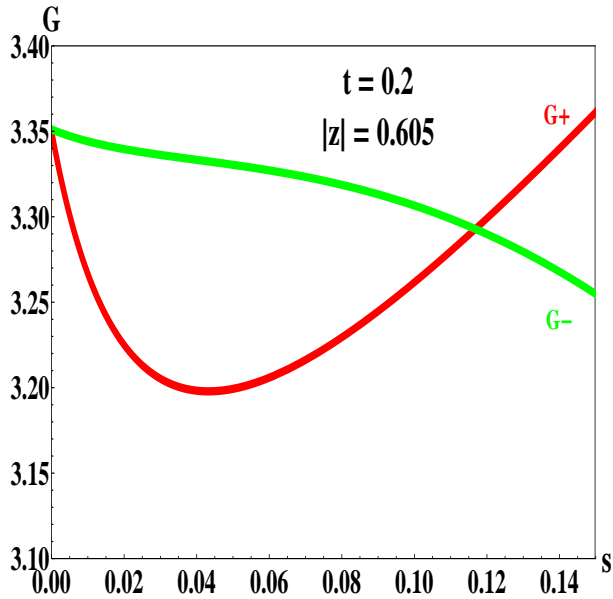}
\end{center}
\caption{\small  Illustrative intersecting point solution plots,
with $G$ versus $s$.
 }
\vspace*{.3in}
\hrule
\label{Gs2}
\end{figure}
\begin{figure}[!t]
\begin{center}
\includegraphics[height=6.0cm,width=7.0cm]{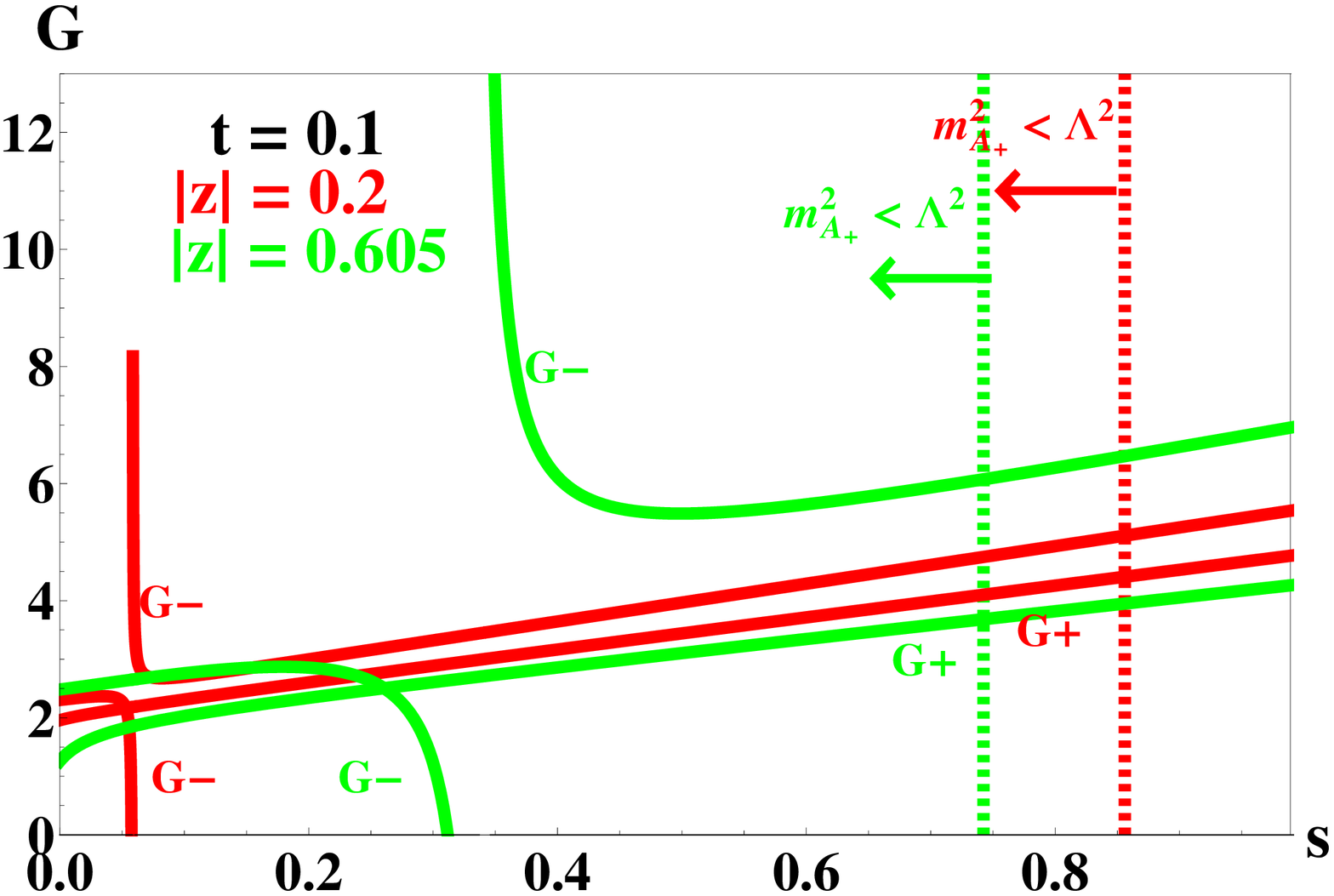}
\hspace*{.1in}
\includegraphics[height=6.0cm,width=7.0cm]{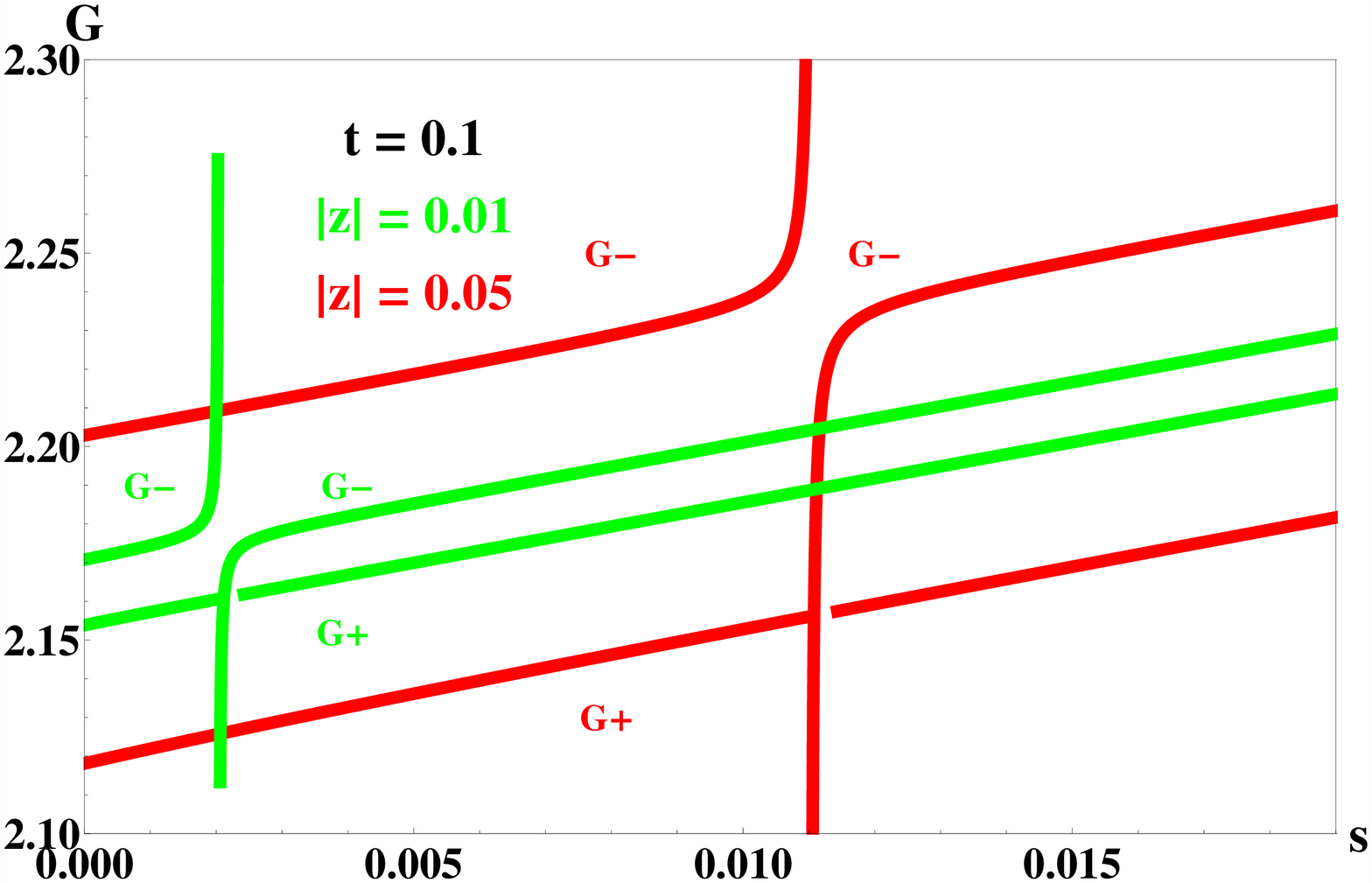}
\end{center}
\caption{\small  Illustrative intersecting point solution plots,
with $G$ versus $s$.; two cases in each frame for comparison.
The two colors each corresponds to the case of one set of
fixed parameter values as shown. Intersecting points of $G^+$
and $G^-$ curves of the same color give the solution point
for the value of $|z|$.
 }
\vspace*{.1in}
\hrule
\label{Gs4}
\end{figure}

We scanned on the effective coupling $G$ versus $s$, $|z|$, and $t$
plots to study the behavior of the intersecting point solutions and check 
for consistence.  The results are as follow:  For somewhat large $t$, solution 
exists only at large enough $|z|$, for example the minimal $|z|$ value for 
solution at $t=.3$ is about 1. Such solution certainly violates (\ref{sb}).  
Actually, solution satisfying the constraint shows up only for $t$ below
about $.265$, which also guarantees the $G_+$ curve to be smooth at least 
within the numerical window of interest. Moreover, the $G$ versus $|z|$
plots for any $t$ and $s$ essentially always give two solutions for (nonzero) 
$|z|$. The larger value $|z|$ solution may not even correspond to a larger 
coupling $G$, as shown in Fig.~\ref{Gz}. Also, a $G$ value smaller than the
$|z|=0$ solution is typical. Another illustration of the same coupling value 
issue is given in Fig.~\ref{Gs2} in which we show $G$ versus $s$ plots with 
two intersecting point, particularly including one with $s=0$. Such nonzero 
$|z|$ with $s=0$ solutions are not available for $t$ less than about $.17$. 
For the latter case, the $G$ versus $s$ plots give a single intersecting point. 
In Fig.~\ref{Gs4}, we show comparisons of the intersecting point solutions 
at the same $t$. We have again solution with larger values of the parameter,
$s$ and $|z|$,  for the masses generated corresponding again to smaller
coupling $G$. Recall the standard, obviously physical sensible, solution 
features of the NJL-type model which our $|z|=0$ solutions shown above 
bears, is that nontrivial symmetry breaking mass solution exists for large 
enough coupling beyond a minimal critical value and increases with the coupling.
The $|z| \ne 0$ `solutions' behaves, however, in ways difficult to understand. 
A more careful inspection of the various plots shows that the $G_-$ curve
in particular has strange singularities. In fact, each intersecting point `solution' 
corresponds to a pair of $s$ and $|z|$ values with the $G_-$ curve either 
diverging at a smaller $s$ or at a smaller $|z|$ value. It sounds like in order 
to `get' to that `solution', one has to bring the coupling value all the way to
positive or negative infinity and back. However, it should be note that nonzero
$\tilde{\eta} (= |z| \sqrt{t})$ increases the mass of one of the smaller mode
but decreases that of the other one [{\em cf.} Eq.(\ref{mA})]. It is not so
trivial to consider if larger $\tilde{\eta}$ or $|z|$ should really be considered
to be giving a larger supersymmetry breaking effect. Another noteworthy
feature is that among solutions of fixed $|z|$  a larger $t$ generally tends 
to give smaller $s$ or $\tilde{m}^2$, and among solutions of fixed $s$,  
a larger $t$ generally tends to give larger $|z|$; larger $t$ always tends to 
increase coupling $G$ required for a solution. Recall that the $|m|$ or $t$ 
value also suppresses the mass generation in the $|z|=0$ case, but $|m|=0$ 
gives certainly no  $|z| \ne 0$ solution.

\section{The Goldstino and Composite (Super)field Dynamics}
Some components of the superfield $U$, which are auxiliary as introduced, 
develop kinetic terms through wavefunction renormalizations in the effective 
theory below the cutoff $\Lambda$. We trace them here through checking of
the relevant loop diagrams, based on the effective Lagrangian in terms of 
components of $\Phi_{\!\ssc R}$ and couplings all having the $\Phi$ 
wavefunction renormalization from the gap equation result incorporated 
[{\em cf.} equations in Appendix A]. The analysis focuses on results at the 
supersymmetry breaking vacuum solutions, {\em i.e.} nonzero $\tilde{m}^2$
with zero or nonzero $\tilde{\eta}$. We only sketch the key results here, 
leaving some more details in Appendix A.

We start with the two-spinors $\chi$ and $\lambda$. The chirality
conserving part of the self-energy diagrams give rise to kinetic terms.
We can see all terms are nonzero in the presence of nonvanishing $\tilde\eta$, 
while the $\chi$-$\lambda$ kinetic mixing vanishes at $\tilde\eta=0$. Full results
are presented in Appendix A. To look at the mass values is complicated. One 
needs first to take a unitary transformation on the hermitian matrix and kinetic 
terms to diagonalize it.  Denote the eigenvalues by $N_{\!f_1}$ and $N_{\!f_2}$, 
and the diagonalizing matrix by $T$. The canonically normalized fermionic modes
are given by
\bea
  \left( \begin{array}{c} f_{\!\ssc 1}  \\ f_{\!\ssc 2}   \end{array}\right)
= \left( \begin{array}{cc}
\frac{1}{\sqrt{N_{\!f_1}}}& 0 \\
                                                  0 & \frac{1}{\sqrt{N_{\!f_2}}}
\end{array}\right)     T
 \left( \begin{array}{c} \chi  \\ \lambda \end{array}\right) \;.
\eea
Only the mass matrix for the canonically modes can be diagonalized
to give the mass eigenvalues. The  mass matrix $\mathcal{M}_f$ for
$ f_{\!\ssc 1}$ and  $f_{\!\ssc 2}$ is hence given by
\bea
 \mathcal{M}_f =
 \left( \begin{array}{cc}
{\sqrt{N_{\!f_1}}}& 0 \\
     0 & {\sqrt{N_{\!f_2}}}
\end{array}\right)      T
 \left( \begin{array}{cc}
\mathcal{M}_{\chi\lambda}
\end{array}\right)
 T^{\ssc T}    \left( \begin{array}{cc}
{\sqrt{N_{\!f_1}}}& 0 \\
    0 & {\sqrt{N_{\!f_2}}}
\end{array}\right)     \;,
\eea
where $\mathcal{M}_{\chi\lambda}=
\left( \begin{array}{cc}
0 & \mu\\
 \mu &    0
\end{array}\right)   +\Omega$, the first part being the tree-level
mass while the last is the matrix for chirality-flipping pieces of
self-energy diagrams.
We have
\bea
\det{\mathcal{M}_f }=N_{\!f_1} N_{\!f_2} \det{\mathcal{M}_{\chi\lambda}} \;.
\eea
\begin{figure}[!t]
\begin{center}
\includegraphics[scale=1]{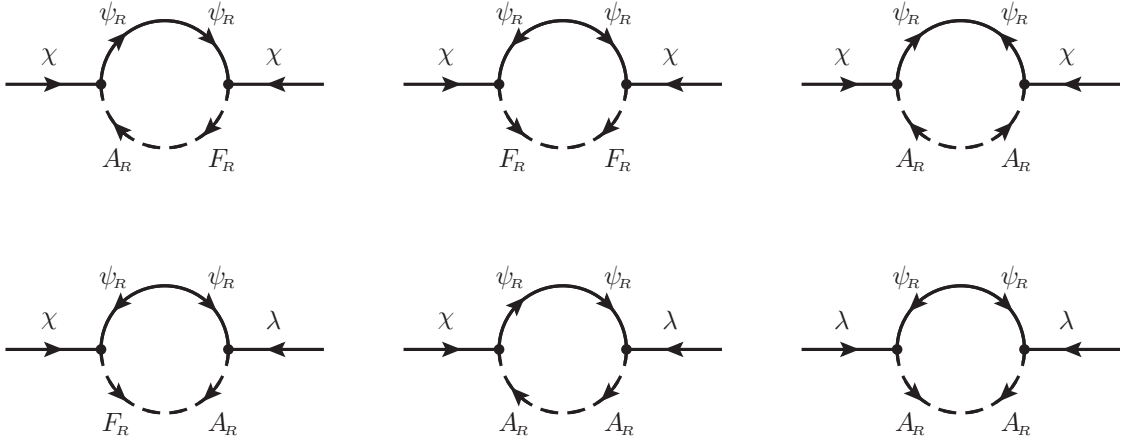}
\end{center}
\caption{\small  Diagrams for fermion masses.}
\vspace*{.1in}
\hrule
\label{fm}
\end{figure}
In the case that the matrix of kinetic terms has the full rank, a zero determinant 
of $\det{\mathcal{M}_f }$ or equivalently $\det{\mathcal{M}_{\chi\lambda}}$
shows the existence of a Goldstino, which is to be expected from the
supersymmetry breaking. We are here mostly interested only in the kind of 
qualitative questions here, which saves us from having the deal with the  
diagonalization of the matrix of kinetic.  For the chirality-flipping diagrams 
(see Fig.~\ref{fm}), dropping the $p$-dependent parts, we have the mass 
terms
\bea
\Omega_{\ssc\chi\chi} &=&
-  \frac{g^2\tilde{m}^4}{\tilde{\eta}}|m|^2 
I_{\!\ssc 3\!F}(|m|^2 ,m_{\!\ssc A_-}^2,m_{\!\ssc A_+}^2) 
+ \frac{1}{2\tilde{\eta}}  \left( g^2 I_{\!\ssc C} + \tilde{m}^2  g^2 I_{\!\ssc N'} \right) \;,
\nonumber \\
\Omega_{\ssc\chi\lambda} &=&
2 \mu g^2 \tilde{m}^2 |m|^2
I_{\!\ssc 3\!F}(|m|^2 ,m_{\!\ssc A_-}^2,m_{\!\ssc A_+}^2)
 -\mu   g^2 I_{\!\ssc N'}  \;,
\nonumber \\
\Omega_{\ssc\lambda\lambda} &=& -\mu^2g^2 \tilde{\eta}	|m|^2  
I_{\!\ssc 3\!F}(|m|^2 ,m_{\!\ssc A_-}^2,m_{\!\ssc A_+}^2) \;,
\eea
where $I_{\!\ssc 3\!F}(|m|^2 ,m_{\!\ssc A_-}^2,m_{\!\ssc A_+}^2) $ is the 
integral of the product of three Feynman propagators with the mass-squares 
as specified, and we have expressed the results with the $I_{\!\ssc C}$ and 
$I_{\!\ssc N'}$ integrals of the gap equations [{\em cf.} Eqs.(\ref{gIc}) and 
(\ref{gIn'})]. Applications of the gap equations kills the term with the then 
vanishing $\left( g^2 I_{\!\ssc C} + \tilde{m}^2  g^2 I_{\!\ssc N'} \right)$ factor and 
has the $ -\mu   g^2 I_{\!\ssc N'}$ term canceling the tree-level term in the mass 
matrix $\mathcal{M}_{\chi\lambda}$ the determinant of which is then exactly 
zero. Hence, we have established the existence of a Goldstino mode for the 
supersymmetry breaking solution with $\tilde{\eta}\ne 0$. For the $\tilde{\eta}= 0$ 
case, only the off-diagonal term is nonzero, which is a result one can see even 
simply from the $U(1)_{\!\ssc R}$ symmetry considerations. The latter has been 
presented in \cite{062}, with the result  that the tree-level Dirac mass is exactly 
canceled by $\Omega$-matrix upon application of the corresponding gap equation 
giving $\mathcal{M}_{\chi\lambda}$ as the zero matrix. Again, we have the
Goldstino. Supersymmetry is really a local/spacetime symmetry. The Goldstino 
would eaten up by the gravitino which would than be massive.

\begin{figure}[!t]
\begin{center}
\includegraphics[scale=1]{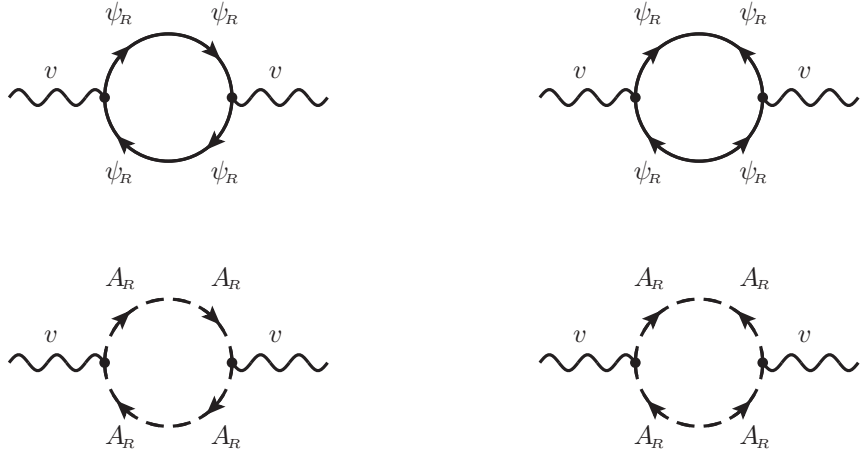}%
\end{center}
\caption{\small  Proper self-energy diagrams for the spin one composite $v^\mu$.}
\vspace*{.1in}
\hrule
\label{v2}
\end{figure}

The spin one vector boson $v^\mu$  is an important characteristic of the model. 
The proper self-energy diagrams (see Fig.~\ref{v2}) for the vector mode give 
the result 
\bea
\frac{-1}{2}\Sigma_{v} &=&
 p^2 \; \frac{\mu^2 g^2}{8}   \bigg\{   12  I_{\!\ssc 2\!F}(|m|^2,|m|^2)
  - 40 |m|^2  I_{\!\ssc 3\!F}(|m|^2,|m|^2,|m|^2)
\nonumber\\  && \quad
   + 32 |m|^2  I_{\!\ssc 4\!F}(|m|^2,|m|^2,|m|^2,|m|^2) 
  +  3  I_{\!\ssc 2\!F}(m_{\!\ssc A_-}^2,m_{\!\ssc A_-}^2)
    - I_{\!\ssc 2\!F}(m_{\!\ssc A_-}^2,m_{\!\ssc A_+}^2)
\nonumber\\  && \quad
   + 3  I_{\!\ssc 2\!F}(m_{\!\ssc A_+}^2,m_{\!\ssc A_+}^2)
  - 4 m_{\!\ssc A_-}^2  I_{\!\ssc 3\!F}(m_{\!\ssc A_-}^2,m_{\!\ssc A_-}^2,m_{\!\ssc A_-}^2)
  - 4 m_{\!\ssc A_+}^2  I_{\!\ssc 3\!F}(m_{\!\ssc A_+}^2,m_{\!\ssc A_+}^2,m_{\!\ssc A_+}^2)
 \nonumber\\  && \quad
- \left( m_{\!\ssc A_-}^2 +m_{\!\ssc A_+}^2 \right)
\left[  I_{\!\ssc 34}(m_{\!\ssc A_-}^2,m_{\!\ssc A_+}^2)
 + I_{\!\ssc 34}(m_{\!\ssc A_-}^2,m_{\!\ssc A_+}^2) \right]
\bigg\}
\nonumber\\  && \quad
+ \frac{\mu^2 g^2}{4} \bigg\{2  I_{\!\ssc F}( |m|^2)
 + I_{\!\ssc F}(m_{\!\ssc A_-}^2)+  I_{\!\ssc F}(m_{\!\ssc A_+}^2)
 -4 |m|^2 I_{\!\ssc F}(|m|^2,|m|^2)
\nonumber\\  && \qquad\quad
 - \left( m_{\!\ssc A_-}^2 + m_{\!\ssc A_+}^2 \right)
   I_{\!\ssc 2\!F}(m_{\!\ssc A_-}^2,m_{\!\ssc A_+}^2)  \bigg\}+ \cdots \;,
\nonumber \\ && _{\tilde{\eta}=0}\longrightarrow
 p^2 \; \frac{\mu^2 g^2}{8}   \Big[ 12  I_{\!\ssc 2\!F}(|m|^2,|m|^2)
  +  5  I_{\!\ssc 2\!F}(m_{\!\ssc A}^2,m_{\!\ssc A}^2)
  - 40 |m|^2  I_{\!\ssc 3\!F}(|m|^2,|m|^2,|m|^2)
\nonumber\\  && \qquad\qquad \qquad
 - 20 m_{\!\ssc A}^2  I_{\!\ssc 3\!F}(m_{\!\ssc A}^2,m_{\!\ssc A}^2,m_{\!\ssc A}^2)
   + 32 |m|^2  I_{\!\ssc 4\!F}(|m|^2,|m|^2,|m|^2,|m|^2)
\nonumber\\  && \qquad\qquad \qquad
 +16  I_{\!\ssc 4\!F}(m_{\!\ssc A}^2,m_{\!\ssc A}^2,m_{\!\ssc A}^2,m_{\!\ssc A}^2 )   \Big]
\nonumber\\  && \qquad\qquad
+ \frac{\mu^2 g^2}{2} \Big[    I_{\!\ssc F}( |m|^2)
 +  I_{\!\ssc F}(m_{\!\ssc A}^2)  -2 |m|^2 I_{\!\ssc F}(|m|^2,|m|^2)
 -  m_{\!\ssc A}^2  I_{\!\ssc 2\!F}(m_{\!\ssc A}^2,m_{\!\ssc A}^2)    \Big] ,
\eea
with $I_{\!n\ssc F}$ denoting the integrals with product of $n$ Feynman propagators.
There is also a tree-level mass-square of $\mu^2$ to be added. It sure indicates that we 
have properly behaved kinetic and mass terms generated (note our metric convention).

The other scalar modes acquire also kinetic and mass terms accordingly.  Mode 
mixings, however, make the result a lot less transparent. Details are given in
Appendix A.

\section{Some Discussion about the Unconventional Features and
the Vacuum Solutions}
The model we have here is quite an unconventional one in many aspects, and 
hence has behavior different from most of the conventional models to the 
extent that many `generic' features of superfield theory or theory with 
spontaneous supersymmetry breaking are simply not present. That may make 
some readers uncomfortable or suspicious.  Hence, we want to address the
unconventional features directly here as much as we can, in relation to the 
validity of our main results of the supersymmetry breaking vacuum solution.

First of all, the basic model Lagrangian is unconventional. It has a four-superfield 
term with like the `wrong' sign and unusual color index contraction, say in 
comparison to the old SNJL model. The color index contraction gives the NJL-type
composite $U$ as an unconventional superfield analysis of which is difficult to 
find in the literature. We will look at that in more details below. Actually, we are 
not the first to write down a quartic term in the K\"ahler potential with a negative 
sign \cite{KS}. To see better its unconventional feature, let us take a look at the 
component field picture of the model. For simplicity, we again drop the color 
index from our analysis. The Lagrangian is given by
\bea  {\cal L} &=&
 i\partial_{\mu}\bar{\psi}\bar\sigma^{\mu}\psi
  + \partial^{\mu}A^*\partial_{\mu}A  +  F^*F \,
+ \left(m AF  - \frac{m}{2}\psi\psi +  h.c. \right)
- \frac{g^2}{2} | 2 F A -\psi \psi |^2
\nonumber\\&&
 + 2 g^2 A^*A\, \partial^{\mu}A^* \partial_{\mu}A
 - 2 g^2  i\partial_{\mu}\bar{\psi}\bar\sigma^{\mu}\psi   A^*A
 - 2 g^2  i \bar{\psi} \bar\sigma^{\mu} \psi A \partial_{\mu}A^* \,.
\eea
From the equation of motion for the auxiliary field $F^*$, we have
\beq \label{F}
F = - \frac{ (m^*  + g^2 \psi\psi)  A^*}{1-2g^2 |A|^2 }\;.
\eeq
The somewhat complicated fractional form of $F$ indicates that the 
component field Lagrangian with $F$ eliminated would have less than 
conventional interaction terms. Naively, the scalar potential is given by
\beq
-V_s = F^*F - 2g^2 A^*A F^*F + m A F + m^* A^* F^* \;.
\eeq
Eliminating $F$ gives, however,
\beq
V_s= \frac{(|m|^2 - g^4 \psi\psi\bar{\psi}\bar{\psi}) |A|^2}{1-2g^2 |A|^2 }\;
\eeq
which formally no longer involves only the scalar. It is suggestive of a
bifermion condensate which fits in the general picture of the NJL setting.
It is interesting to note that for  $m=0$ the model actually has no pure scalar
part in $V_s$, for any coupling $g^2$. On the other hand, if one neglects
the fermion field part in the above, the potential looks simple enough,
$V_s= \frac{|m|^2 |A|^2}{1-2g^2 |A|^2 }$, with a supersymmetric minimum
at zero $A$. For positive $g^2$, however, it blows up at  $|A|=1/\sqrt{2}g$.
For a perturbative coupling, one expect $1/g$ bigger than the model
cut-off scale $\Lambda$, hence the potential is well behaved within
the cut-off. With strong coupling $g^2$, one cannot be so comfortable.
In fact, for $|A|>1/\sqrt{2}g$, the potential goes negative, contradicting
our expectation for a supersymmetric model. The analysis so far suggests
compatibility with a plausible nonperturbative supersymmetry
breaking. In fact, the analysis here illustrates clearly that for strong
enough coupling the model has no sensible perturbative vacuum, not
even the naive supersymmetric vacuum one may naively expect to work,
at least in the $m=0$ case or for large $g^2$.

The nonperturbative NJL-type feature is what gives the model a sensible
vacuum. In fact, the model other than being a superfield one has mostly
quite conventional NJL-type features at least for the $\tilde{m}^2\ne 0$
and $\tilde{\eta}=0$ vacuum.  Here below, we mostly address only
the latter case as our supersymmetry breaking solution. Let us now take
a look at the scalar potential in the presence of the composite $U$,
namely as described by the effective theory Lagrangian, for $m=0$ at
the tree-level. We have to emphasize that the effective theory really
comes from the NJL-type composite (super)field thinking consistence of
which asks for the potential analysis as performed above in Sec. III. We
are looking at the tree-level potential here only to illustrate further the
unconventional features of the model, here as given by the effective theory
Lagrangian.  The potential has the very unconventional form given by
\bea
V_{e\!f\!\!f}^{\mbox{\tiny tree}}
= -F_{\!\ssc R}^*F_{\!\ssc R} -g CF_{\!\ssc R}^*F_{\!\ssc R}
-\mu CD-N^*N-\mu g D A_{\!\ssc R}^*A_{\!\ssc R}
-g NA_{\!\ssc R}F_{\!\ssc R}^* - gN^*A_{\!\ssc R}^*F_{\!\ssc R} \;,
\eea
which has vanishing minima both at the trivial supersymmetric
origin of the field space except with arbitrary value for
$A_{\!\ssc R}$, and at $C=D=0$ and $|A_{\!\ssc R}|=1/g$ with
$F_{\!\ssc R}$ and $N=-gA_{\!\ssc R}^*F_{\!\ssc R}$ undetermined. The
latter potentially supersymmetry breaking minimum is  fully consistent
with expressions one may obtain naively for eliminating any of the auxiliary 
scalars including $F$.  Taking $m=0$ simplifies the analysis.  It is also really 
the most interesting benchmark case in which there is no input mass 
parameter at all. Let us emphasize again that the composite superfield $U$ 
though has a spin one component $v^\mu$ is not at all the conventional real 
scalar superfield of a  gauge boson. Before putting in the `quark-loop' correction,
it has no kinetic term and the conventional $D^2$ term is missing. It  is sure 
massive, and does not even have the right couplings to be considered as the 
gauge boson with broken gauge symmetry. 

Eliminating all $C$, $N$, and $D$ from $V_{e\!f\!\!f}^{\mbox{\tiny tree}} $
of course gives back only the $V_s$ potential of the origin Lagrangian.
Those conditions are really from the composite condition of
$U=-\frac{g}{\mu} \bar{\Phi}_{\!\ssc R}{\Phi}_{\!\ssc R}$ the NJL
wisdom of nontrivial two field condensates says exactly that they
should not be applied to the VEVs.  A simple conclusion here is that while 
our gap equations above allows a supersymmetry preserving solution, the 
conventional wisdom of that being the preferred vacuum may not apply. At 
least there is no indications that the supersymmetry breaking vacuum is less 
stable. In fact, the supersymmetric solution means there is no two-field
condensate of $\Phi$ except possibly what contributes to $\lla C\rra$. The 
latter is just a wavefunction renormalization factor for $\Phi$ which is 
already absorbed in $V_{e\!f\!\!f}^{\mbox{\tiny tree}}$ and gives the same 
form for $V_s$ in terms of renormalized quantities. After all, without 
symmetry breaking two-field condensate is like what one expect with weak  
$g^2$ coupling, there should not be any composite field degree of freedom 
and the perturbative tree-level $V_s$ should be expected to give the correct 
qualitative feature of the solution.  However, the latter looks unstable or
sick for strong enough coupling.

Another point to note is that the gap equation analysis for an NJL-type model 
always seems to admit the symmetry preserving solution \cite{BE,042,050}.  
However, we can see that the symmetry preserving solution looks applicable 
only to the case of weak $g^2$, or subcritical $G$ coupling. The derivations 
of the gap equation(s) have taken no constraint on the coupling. Nontrivial 
solution is not possible with a subcritical coupling. Hence, admitting the trivial
solution is really a like consistence condition. Now if one assumes the composite 
formation and looks at the effective theory,  the following argument says the 
trivial, symmetry preserving solution should not be considered valid. At least
 in principle, we can take the effective theory with the `quark-loop' contributions
 included without putting in any symmetry breaking VEVs. That has been done 
for the NJL and the SNJL models, for example, to retrieve effective potential for 
the SM and the (M)SSM \cite{C,CLB}. Those scalar potentials do not admit the 
symmetry preserving vacuum solutions. We want to emphasize again that the 
effective theory picture should be taken as valid only with the strong coupling.
Note further that the gap equations as obtained from the tadpole analysis of the 
potential for the effective theory formally gives only vanishing first derivative 
conditions, hence turning points rather than sure minimum. For the SM or the 
(M)SSM, the zero VEV symmetry preserving point for the Higgs potential is sure 
a turning point, and indeed a local maximum. In that light, the nature of the 
supersymmetry preserving point as a vacuum solution may be questionable 
at large coupling. That thinking is in consistent with the $V_s$ analysis. 

From all the above,  we conclude that there is no clear indication of the 
supersymmetry breaking vacua being metastable or not stable enough. It 
actually may be the preferred solution for the case of strong coupling. It will 
of course be very nice if what we argue for here above can be rigorous 
demonstrated to be the case with some further analysis, probably beyond 
the large $N$ approximation as inherent in the basic NJL-type analysis here.
A related and important issue is the relative stability of the different
supersymmetry breaking vacua. While it is difficult to check if the same
strong coupling value admits more than one such vacua, it is seems certain to 
be the case for the coupling values that admit $\tilde{\eta}\ne 0$ solution(s),
as the existence of $\tilde{\eta}= 0$ and $\tilde{m}^2\ne 0$ solution is generic 
once the coupling is beyond the critical value. The problem is for the nonzero
input $m$ case only though. 

Another important aspect about our model that some may feel suspicious is 
its being able to avoid the vanishing supertrace condition for the mass-squares 
of the component fields, which is in general difficult. However, that the condition 
was established only for specific models of supersymetry breaking rather than 
as a generic result \cite{st,Grisaru:1982sr,PBin}, though the class of models
include most of the better known ones. It sure does not work for the case of 
an anomalous Fayet-Iliopoulos $D$-term for example. There is no good reason 
to assume any conclusion on the issue about the case of our  model.  If there 
is at all any somewhat similar structure in our model to those more conventional
ones in terms of supersymmetry breaking, it may be with the Fayet-Iliopoulos 
case with a potential sort of linear in the $D$-term. And we sure have no anomaly 
issue.  Our analysis clearly indicates generation of only soft supersymmetry 
breaking masses for the scalar component $A$ and not  the fermion $\psi$. There
cannot be any doubt about that. In fact, the $m=0$ case with $\tilde{m}^2\ne  0$ 
and $\tilde{\eta}=0$ solution is even more illustrative. There are clearly masses 
for $A$ and the composite spin-one $v^\mu$ generated, while no mass term for 
any fermionic mode at all.  Avoiding the vanishing supertrace condition of course
is the central feature that we are after, with particularly interesting application 
phenomenologically.

\section{Remarks and Conclusions}
Dynamical supersymmetry breaking is an interesting and important topic 
\cite{SS}. Our new simple model with a single chiral  superfield (multiplet) 
should be a great addition to the latter. The fact that the model has as its 
direct consequence the generation of soft supersymmetry breaking masses 
is specially interesting in view of the requirement of soft supersymmetry 
breaking in any low energy phenomenological application of supersymmetry 
as in a SSM.  We want to emphasize that our key interest is really the case 
with an $SU(N)$ multiplet and no input mass parameter, {\em i.e.} $m=0$. 
For the case, the analysis can simply be considered one with the color index 
hidden so long as we put back the  color factor $N$ in the relevant loop 
diagrams, basically has the $g^2$ factor in all those results including the gap 
equations to be replaced by $g^2 N$. We may also take the superfield as like 
one of the chiral matter superfield multiplets in the SSM. Together with our 
earlier HSNJL model \cite{034,042}, a simple SSM with all (super)symmetry 
breaking and mass parameters generated dynamically is easily in sight, 
though it remains to see if a model with only the SSM superfield spectrum 
minus the Higgs supermultiplets can be a consistent model theoretically and
phenomenologically. Even if the answer to the latter question is a no, it looks 
like there is at least enough room to have a model with like a single extra chiral 
superfield to produce the supersymmetry breaking. We consider a model of 
such kind quite compelling as an alternative to the full models of the SSM in 
the literature having the extra sectors. The current study is a big step in the
direction, to which we sure love to further our investigations. The model 
mechanism of course may also be applied to other model building works, 
for example in addressing the (S)SM flavor structure questions \cite{059}.

We take only the case of a simple singlet composite of
$U \sim \Phi^\dag_a \Phi^a$ here. A somewhat more complicated
case as studied in the case of (non-supersymmetric) NJL-type composite of spin 
one field \cite{S} would have the composite in the adjoint representation. 
Similar but superfield version of four-superfield interactions may be considered 
though not in relation to pure soft supersymmetry breaking. It is also possible 
to have a model in which the composite superfield  $U$ behaves like a massive
gauge field supermultiplet \cite{064}, much in parallel with the
non-supersymmetric models of Ref.\cite{S}. Note again that our current model 
does not have the right coupling for the spin one field $v_\mu$ to behave like 
a gauge boson at all. There is no $A^*v^\mu v_\mu A$ term in the Lagrangian 
[{\em cf.} Eq.(\ref{Lec}) and Eq.(\ref{Lecq})], or no $\bar{\Phi} U^2 \Phi$ term 
Eq.(\ref{Le}) in the superfield picture. It is possible to think about the 
electroweak gauge bosons as such composites. However, we echo the author 
of  Ref.\cite{S} against advocating the kind of scenario.

Finally, we emphasize that with the modern effective (field) theory perspective, 
it is the most natural thing to consider any theory as an effective description of 
Nature only within a limited domain/scale. Physics is arguably only about 
effective theories, as any theory can only be verified experimentally up to a 
finite scale and there may always be a cut-off beyond that. Having a cutoff scale 
with the so-called nonrenormalizable higher dimensional operators is hence in 
no sense an undesirable feature. Model content not admitting any other 
parameter with mass dimension in the Lagrangian would be very natural. 
Dynamical mass generation with symmetry breaking is then necessary to give 
the usual kind of low energy phenomenology such as the Standard Model one. 
That is actually the key motivation behind our  line of work on the subject matter.

\bigskip\bigskip

\appendix
\section{Some more technical details of the model calculations}
Starting with the effective Lagrangian of Eq.(\ref{Lec}) with $c$, $d$,
and $n$ denoting VEVs of the (original) scalars $C$, $D$, and $N$, we have,
in terms of the renormalized components $A_{\!\ssc R}=\sqrt{Z} A$,
$\psi_{\!\ssc R}=\sqrt{Z} \psi$, and $F_{\!\ssc R}=\sqrt{Z} F$ of
$\Phi_{\!\ssc R}=\sqrt{Z} \Phi$ with the common (supersymmetric) wavefunction renormalization factor $Z=1+g_o c$, the quadratic part of the Lagrangian is given by
\bea
{\cal L}_{ef\!f}^{(2)}
&=&
          A_{\!\ssc R}^*\Box A_{\!\ssc R}
         +i(\partial_\mu\bar{\psi}_{\!\ssc R}) \bar{\sigma}^\mu\psi_{\!\ssc R}
         + F_{\!\ssc R}^* F_{\!\ssc R}   
+\frac{m}{2}   \left( 2 A_{\!\ssc R} F_{\!\ssc R}  -\psi_{\!\ssc R}\psi_{\!\ssc R}  \right)
+\frac{m^*}{2}  \left(
   2 A_{\!\ssc R}^* F_{\!\ssc R}^*  -\bar{\psi_{\!\ssc R}}\bar{\psi_{\!\ssc R}} \right)
\nonumber\\  &+&
 {\mu}CD  -{\mu}\chi\lambda -{\mu}\bar{\chi}\bar\lambda
+N\bar{N} -\frac{\mu^2}{2}v^{\nu}v_{\nu}
+ \mu g d   A_{\!\ssc R}^* A_{\!\ssc R}
+g n A_{\!\ssc R} F_{\!\ssc R}^*
+g n^* A_{\!\ssc R}^* F_{\!\ssc R}
\;, \label{Lecq}
\eea
in which we have the renormalized mass and coupling $m=\frac{m_0}{Z}$
and $g=\frac{g_0}{Z}$. Here the scalars $C$, $N$, and $D$ are the physical
ones with VEVs already pull out, though we do not distinguish them from
the original ones with VEVs explicitly in notation. One can easily obtain
the following propagator expressions :
\bea
\langle T(A_{\!\ssc R}\,A_{\!\ssc R}^*)\rangle
&=&\frac{-i ( p^2+ |m|^2 + g^2 |n|^2 -\mu g d)}
   { ( p^2+ |m|^2 + g^2 |n|^2 -\mu g d)^2 - 4 g^2 |n|^2 |m|^2  }  \;,
\nonumber\\
\langle T(A_{\!\ssc R}\,A_{\!\ssc R})\rangle
&=&\frac{ 2i g n^* m^* }
   { ( p^2+ |m|^2 + g^2 |n|^2 -\mu g d)^2 - 4 g^2 |n|^2 |m|^2  }  \;,
\nonumber\\
\langle T(F_{\!\ssc R}\,F_{\!\ssc R}^*)\rangle
&=&\frac{i (p^2  -\mu g d) ( p^2+ |m|^2 + g^2 |n|^2 -\mu g d)}
   { ( p^2+ |m|^2 + g^2 |n|^2 -\mu g d)^2 - 4 g^2 |n|^2 |m|^2  }    \;,
\nonumber\\
\langle T(F_{\!\ssc R}\,F_{\!\ssc R})\rangle
&=&\frac{ - 2i g n m^* (p^2  -\mu g d)}
   { ( p^2+ |m|^2 + g^2 |n|^2 -\mu g d)^2 - 4 g^2 |n|^2 |m|^2  }    \;,
\nonumber\\
\langle T(A_{\!\ssc R}F_{\!\ssc R})\rangle
&=&\frac{i m^*   ( p^2+ |m|^2 - g^2 |n|^2 -\mu g d)}
   { ( p^2+ |m|^2 + g^2 |n|^2 -\mu g d)^2 - 4 g^2 |n|^2 |m|^2  }   \;,
\nonumber\\
\langle T(A_{\!\ssc R}F_{\!\ssc R}^*)\rangle
&=&\frac{i g n^*   ( p^2- |m|^2 + g^2 |n|^2 -\mu g d)}
   { ( p^2+ |m|^2 + g^2 |n|^2 -\mu g d)^2 - 4 g^2 |n|^2 |m|^2  }   \;,
\nonumber\\
\langle T(\psi_{{\!\ssc R}_{\alpha}} \bar\psi_{{\!\ssc R}_{\dot\beta}})\rangle
&=& \frac {- i p_{\mu}\,\sigma^{\mu}_{\alpha \dot \beta}}
   {  p^2+|m|^2 }    \;,
\nonumber\\
\langle T(\psi_{{\!\ssc R}_{\alpha}} \psi_{\!\ssc R}^{\beta})\rangle
&=& \frac {-i m^* \delta_\alpha^\beta }
   {  p^2+|m|^2 }    \;.
\label{prop}\eea
Note that $-\mu g d$  and $-gn$ here correspond to the (renormalized) soft
mass terms $\tilde{m}^2$ and $\tilde{\eta}$. The propagator expressions can be
matched to that of the superfield $\Phi$ in Eqn.(\ref{sprop}).

The remaining, interaction, terms in the effective Lagrangian read
\bea
{\cal L}_{ef\!f}^{int}
&=&
g C\left[
          A_{\!\ssc R}^*\Box A_{\!\ssc R}
         +i(\partial_\mu\bar{\psi}_{\!\ssc R}) \bar{\sigma}^\mu\psi_{\!\ssc R}
         + F_{\!\ssc R}^* F_{\!\ssc R}   \right]
-\mu g \psi_{\!\ssc R} \lambda A_{\!\ssc R}^* - \mu g \bar{\psi_{\!\ssc R}}\bar{\lambda} A_{\!\ssc R}
+ \mu g  D  A_{\!\ssc R}^* A_{\!\ssc R}
\nonumber\\  &&
-i\frac{g}{2}\bar{\psi_{\!\ssc R}}\bar{\sigma}^\mu\chi \partial_\mu A_{\!\ssc R}
+i\frac{g}{2}(\partial_\mu\bar{\psi_{\!\ssc R}})\bar{\sigma}^\mu\chi A_{\!\ssc R}
-g\chi\psi_{\!\ssc R} F_{\!\ssc R}^*
+g N A_{\!\ssc R} F_{\!\ssc R}^*
\nonumber\\  &&
+i\frac{g}{2}\bar{\chi}   \bar{\sigma}^\mu\psi_{\!\ssc R}  \partial_\mu A_{\!\ssc R}^*
-i\frac{g}{2} A_{\!\ssc R}^*   \bar{\chi}\bar{\sigma}^\mu \partial_\mu\psi_{\!\ssc R}
-g \bar{\chi}\bar{\psi}_{\!\ssc R}F_{\!\ssc R}
+g N^* A_{\!\ssc R}^* F_{\!\ssc R}
\nonumber\\  &&
-\frac{\mu g}{\sqrt{2}}\eta^{\mu\nu}v_\mu iA_{\!\ssc R}^*\partial_\nu A_{\!\ssc R}
  +\frac{\mu g}{\sqrt{2}}\eta^{\mu\nu}v_\mu i(\partial_\nu A_{\!\ssc R}^*)A_{\!\ssc R}
  -\frac{\mu g}{\sqrt{2}}\eta^{\mu\nu}v_\mu \bar{\psi}_{\!\ssc R}\bar{\sigma}_\nu\psi_{\!\ssc R}
\;. \label{Leci}
\eea
Note that the above gives essentially all parts of the Lagrangian, apart from a 
constant. The linear terms are canceled at the physical vacuum with consistent 
$c$, $n$, $d$ solutions discussed in the main text.

In the following, we present some details of the `quark-loop' contribution to the
two-point functions for the various components of the composite superfield $U$
at the supersymmetry breaking vacuum solutions, as discussed in Sec. V.  Though
we argue in the text that ${\tilde{\eta}\ne 0}$ solution does not look acceptable,
we present fully generic results for completion. The results may offer more
insight into the problem.

\begin{figure}[!t]
\begin{center}
\includegraphics[scale=1]{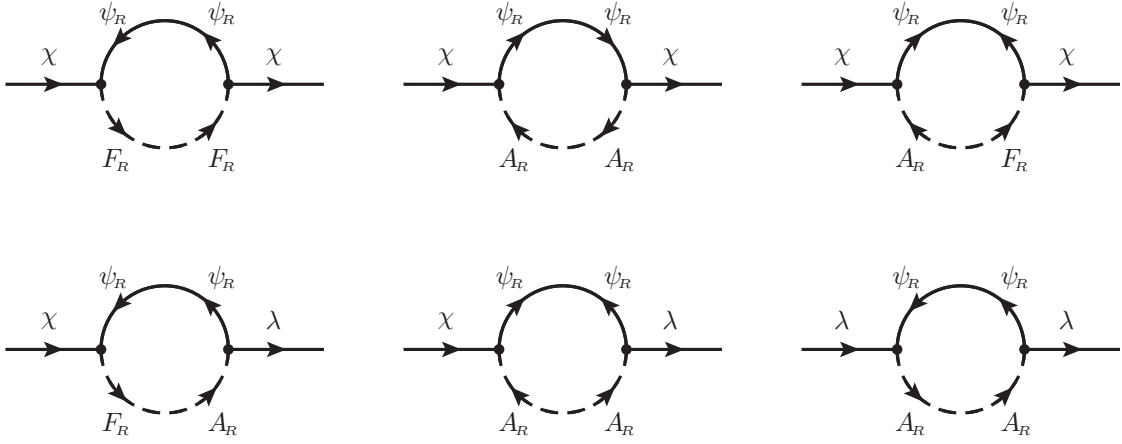}%
\end{center}
\caption{\small  Diagrams for the generation of kinetic terms for the 
fermionic modes.}
\vspace*{.1in}
\hrule
\label{fk}
\end{figure}

The two-point functions for fermion kinetic terms are given by the diagrams
in Fig.~\ref{fk}, with the  $ip\cdot \bar\sigma\Xi$ results given as
\footnote{A good reference for the technical aspects of the
calculations by Ref.\cite{DHM} notation of which we more or less
follow, except that our background notation/convention is
based on Wess and Bagger \cite{WB}.}
\bea
\Xi_{\ssc\chi\chi}
&=& \frac{g^2}{4}
\Big[  |m|\left(|m|-|\tilde{\eta}|\right)  I_{\!\ssc 2\!F}(|m|^2,m_{\!\ssc A_-}^2)
     +|m|\left(|m|+|\tilde{\eta}|\right)  I_{\!\ssc 2\!F}(|m|^2,m_{\!\ssc A_+}^2)
\nonumber\\&&
+2 I_{\!\ssc F}(m_{\!\ssc A_-}^2)
         -2 \left( 2\tilde{m}^2  + 3|m|^2  - 2|\tilde{\eta}||m|\right)
				   I_{\!\ssc 2\!F}(m_{\!\ssc A_-}^2,m_{\!\ssc A_-}^2)
\nonumber\\&&   
    + 2 I_{\!\ssc F}(m_{\!\ssc A_+}^2)
         -2 \left( 2\tilde{m}^2 + 3|m|^2 + 2|\tilde{\eta}||m|\right)
				   I_{\!\ssc 2\!F}(m_{\!\ssc A_+}^2,m_{\!\ssc A_+}^2)
\nonumber\\&&     
  -2|m|^2 \left( m_{\!\ssc A_-}^2-2\tilde{m}^2-3|m|^2+2|\tilde{\eta}||m| \right)
           I_{\!\ssc 3\!F}(|m|^2,m_{\!\ssc A_-}^2,m_{\!\ssc A_-}^2)
\nonumber\\&&    
   -2|m|^2 \left( m_{\!\ssc A_+}^2-2\tilde{m}^2-3|m|^2-2|\tilde{\eta}||m| \right)
           I_{\!\ssc 3\!F}(|m|^2,m_{\!\ssc A_+}^2,m_{\!\ssc A_+}^2)   \Big]+\cdots \;,
\nonumber \\ &&
 _{\tilde{\eta}=0}\longrightarrow    \frac{g^2}{2}
\Big[ 2 I_{\!\ssc F}(m_{\!\ssc A}^2) +  |m|^2 I_{\!\ssc 2\!F}(|m|^2,m_{\!\ssc A}^2)
  -2 \left( 2\tilde{m}^2 + 3|m|^2 \right)    I_{\!\ssc 2\!F}(m_{\!\ssc A}^2,m_{\!\ssc A}^2)
\nonumber \\ && \qquad\qquad\quad
    +2|m|^2 \left( \tilde{m}^2 + 2|m|^2 \right)
           I_{\!\ssc 3\!F}(|m|^2,m_{\!\ssc A}^2,m_{\!\ssc A}^2)   \Big] \;,
\eea
\bea
\Xi_{\ssc\chi\lambda}
&=& 
 \mu g^2\tilde{\eta}^*
\bigg[\left(1-\frac{|m|}{4|\tilde{\eta}|}\right)I_{\!\ssc 2\!F}(|m|^2,m_{\!\ssc A_-}^2)
     +\left(1+\frac{|m|}{4|\tilde{\eta}|}\right)I_{\!\ssc 2\!F}(|m|^2,m_{\!\ssc A_+}^2)\nonumber\\&&   
   -m_{\!\ssc A_-}^2I_{\!\ssc 3\!F}(|m|^2,m_{\!\ssc A_-}^2,m_{\!\ssc A_-}^2)
			  -m_{\!\ssc A_+}^2I_{\!\ssc 3\!F}(|m|^2,m_{\!\ssc A_+}^2,m_{\!\ssc A_+}^2)
\bigg] +\cdots  \;,
\nonumber \\ &&
 _{\tilde{\eta}=0}\longrightarrow  0 \;.
\eea
\bea
\Xi_{\ssc\lambda\lambda}
&=& 
\frac{\mu^2g^2}{2}  \Big[  I_{\!\ssc 2\!F}(m_{\!\ssc A_-}^2,m_{\!\ssc A_-}^2)
     +I_{\!\ssc 2\!F}(m_{\!\ssc A_+}^2,m_{\!\ssc A_+}^2)
\nonumber\\&&	
 -|m|^2I_{\!\ssc 3\!F}(|m|^2,m_{\!\ssc A_-}^2,m_{\!\ssc A_-}^2)
		 -|m|^2I_{\!\ssc 3\!F}(|m|^2,m_{\!\ssc A_+}^2,m_{\!\ssc A_+}^2)  \Big] +\cdots\;,
\nonumber \\ &&
 _{\tilde{\eta}=0}\longrightarrow
2 \mu^2g^2   \left[  I_{\!\ssc 2\!F}(m_{\!\ssc A}^2,m_{\!\ssc A}^2)
 -|m|^2I_{\!\ssc 3\!F}(|m|^2,m_{\!\ssc A}^2,m_{\!\ssc A}^2) \right] \;,
\eea
where $I_{\!n\!\ssc F}$ denote integrals each of a product of $n$ Feynman
propagators with the mass-square parameters as given. We have given besides the 
general result also the simplified expression at the $\tilde{\eta}=0$ limit. Recall
\[
m_{\!\ssc A_\mp}^2=\tilde{m}^2  +(|m|\mp|\tilde{\eta}|)^2
\]
and at the limit $\tilde{\eta}=0$, we have used
$m_{\!\ssc A}^2\equiv\tilde{m}^2  +|m|^2=m_{\!\ssc A_-}^2=m_{\!\ssc A_+}^2$.
The vanishing kinetic mixing between $\chi$ and $\lambda$ can also be easily seen 
from $U(1)_{\!\ssc R}$ symmetry considerations.

\begin{figure}[!b]
\hrule\vspace*{.3in}
\begin{center}
\includegraphics[scale=1]{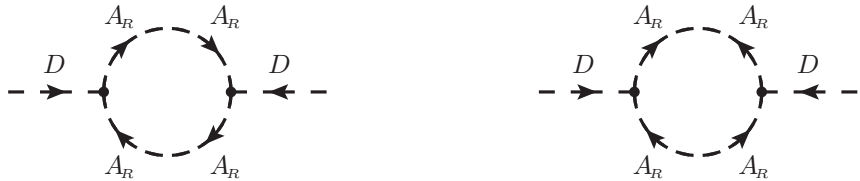}%
\end{center}
\caption{\small  Proper self-energy diagrams for the $DD$ term.}
\vspace*{.1in}
\hrule
\label{dd}
\end{figure}
\begin{figure}[!b]
\hrule
\vspace*{.3in}
\begin{center}
\includegraphics[scale=1]{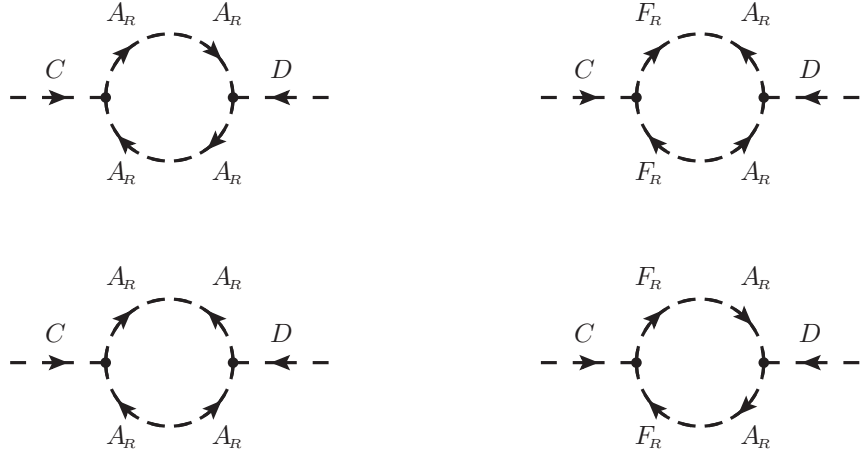}%
\end{center}
\caption{\small  Proper self-energy diagrams for the $CD$ term.}
\vspace*{.1in}
\hrule
\label{cd}
\end{figure}
\begin{figure}[!b]
\begin{center}
\includegraphics[scale=0.8]{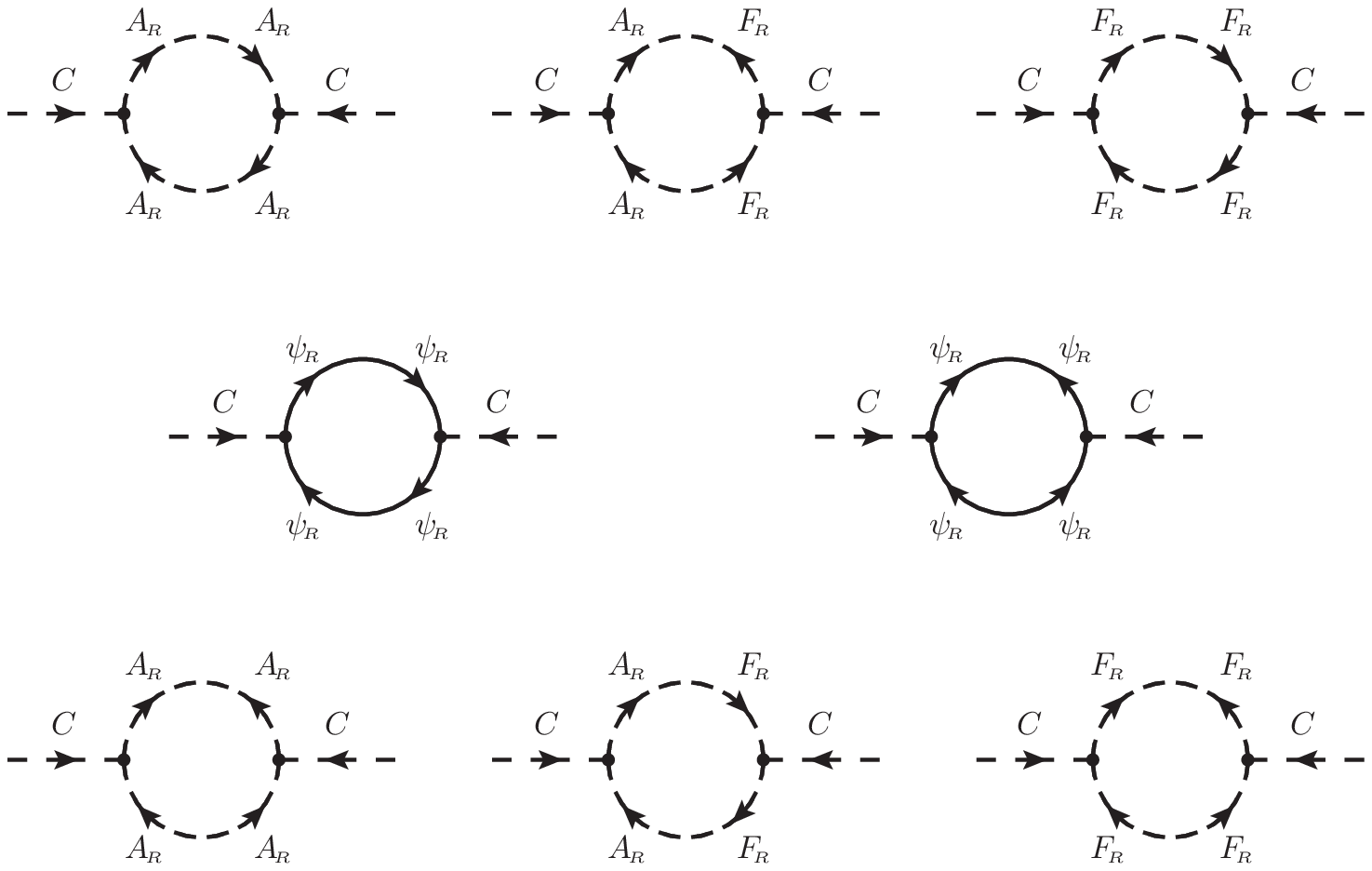}%
\end{center}
\vspace*{-.1in}
\caption{\small  Proper self-energy diagrams for the $CC$ term.}
\hrule
\label{cc}
\end{figure}
\begin{figure}[!t]
\begin{center}
\includegraphics[scale=1]{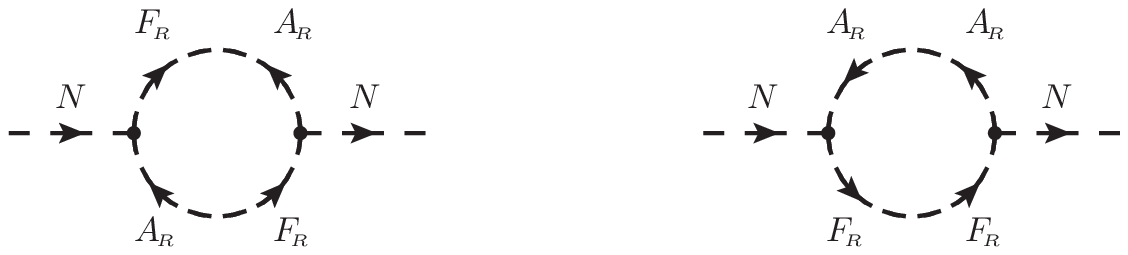}%
\end{center}
\caption{\small  Proper self-energy diagrams for the $NN^*$ term.}
\vspace*{.1in}
\hrule
\label{nnb}
\end{figure}
\begin{figure}[!t]
\vspace*{.1in}
\begin{center}
\includegraphics[scale=1]{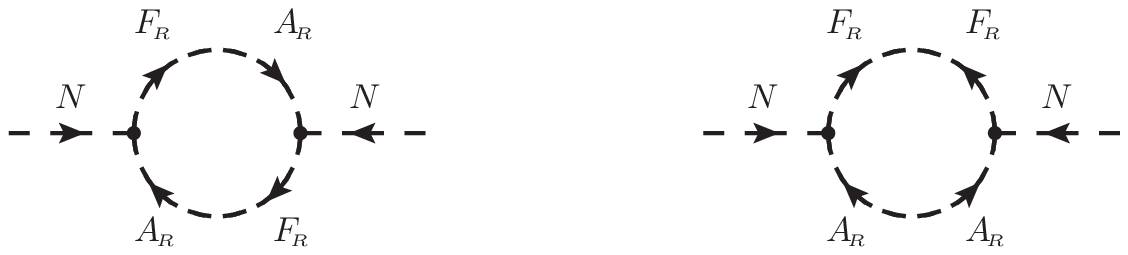}%
\end{center}
\caption{\small  Proper self-energy diagrams for the $NN$ term.}
\vspace*{.1in}
\hrule
\label{nn}
\end{figure}
\begin{figure}[!t]
\begin{center}
\includegraphics[scale=1]{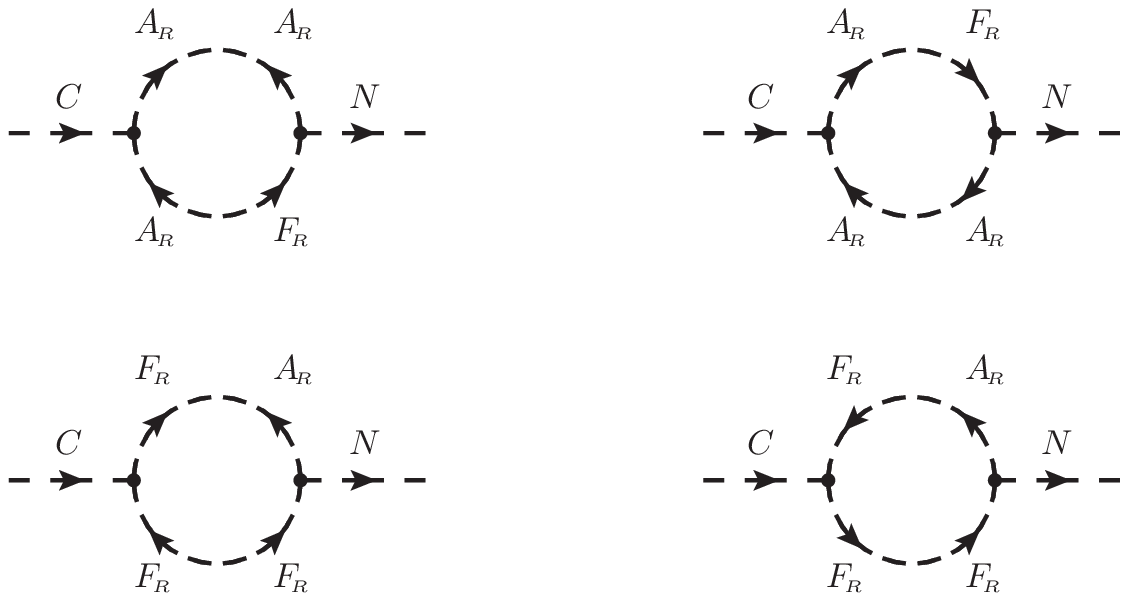}%
\end{center}
\caption{\small  Proper self-energy diagrams for the $CN^*$ term.}
\vspace*{.1in}
\hrule
\label{cn}
\end{figure}
\begin{figure}[!t]
\vspace*{.2in}
\begin{center}
\includegraphics[scale=1]{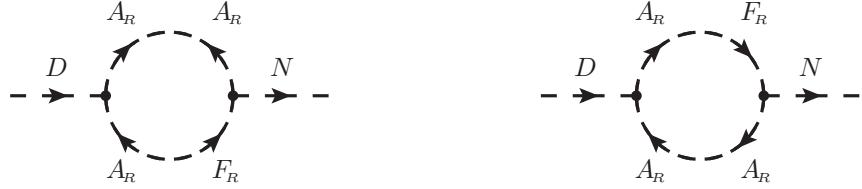}%
\end{center}
\caption{\small  Proper self-energy diagrams for the $DN^*$ term.}
\vspace*{.1in}
\hrule
\label{dn}
\end{figure}

For the scalars, results for the various proper self-energy diagrams are
very tedious. Apart from $I_{\!n\!\ssc F}$, we further introduce
\bea
I_{\!\ssc 34}(m_{a}^2,m_{b}^2) \equiv
  3  I_{\!\ssc 3\!F}(m_{a}^2,m_{b}^2,m_{b}^2)
  - 4 m_{b}^2  I_{\!\ssc 4\!F}(m_{a}^2,m_{b}^2,m_{b}^2,m_{b}^2)\;,
\eea
to present the results. Again, we give the general result and the 
$\tilde{\eta}=0$ limit, and the diagrams are given in the Figs~(\ref{dd}) 
to (\ref{dn}). We have:
\bea
\frac{-1}{2} \Sigma_{\!\ssc D\!D}
\!\! &=& \!\!
 p^2    \frac{\mu^2 g^2}{8}  \big[ I_{\!\ssc 34}(m_{\!\ssc A_-}^2,m_{\!\ssc A_-}^2)
  + I_{\!\ssc 34}(m_{\!\ssc A_+}^2,m_{\!\ssc A_+}^2) \big]
\nonumber \\ && \qquad
 + \frac{\mu^2 g^2}{8}  \big[ 
 I_{\!\ssc 2\!F}(m_{\!\ssc A_-}^2,m_{\!\ssc A_-}^2) +
 I_{\!\ssc 2\!F}(m_{\!\ssc A_+}^2,m_{\!\ssc A_+}^2) \big] 
+ \cdots \;,
\nonumber \\ &&
 _{\tilde{\eta}=0}\longrightarrow
p^2    \frac{\mu^2 g^2}{4} I_{\!\ssc 34}(m_{\!\ssc A}^2,m_{\!\ssc A}^2)
+ \frac{\mu^2 g^2}{4}
 I_{\!\ssc 2\!F}(m_{\!\ssc A}^2,m_{\!\ssc A}^2) \;,
\eea
\bea
\frac{-1}{2} \Sigma_{\!\ssc C\!D}
&=&
p^2 \; \frac{\mu g^2}{4}  \Big\{
\left[ m_{\!\ssc A_-}^2  + \left(|m|-|\tilde{\eta}|\right)^2 \right]
 I_{\!\ssc 34}(m_{\!\ssc A_-}^2,m_{\!\ssc A_-}^2)
 +  \left[ m_{\!\ssc A_+}^2  + \left(|m|+|\tilde{\eta}|\right)^2 \right]
I_{\!\ssc 34}(m_{\!\ssc A_+}^2,m_{\!\ssc A_+}^2)
\nonumber \\  && \; 
+ \frac{\mu g^2}{4} \Big\{
 \left[ m_{\!\ssc A_-}^2  + \left(|m|-|\tilde{\eta}|\right)^2 \right]
   I_{\!\ssc 2\!F}(m_{\!\ssc A_-}^2,m_{\!\ssc A_-}^2)
+  \left[ m_{\!\ssc A_+}^2  + \left(|m|+|\tilde{\eta}|\right)^2 \right]
 I_{\!\ssc 2\!F}(m_{\!\ssc A_+}^2,m_{\!\ssc A_+}^2)
\nonumber \\ && \qquad\qquad
   - I_{\!\ssc F}(m_{\!\ssc A_-}^2) - I_{\!\ssc F}(m_{\!\ssc A_+}^2) \Big\}
 + \cdots \;,
\nonumber \\ && _{\tilde{\eta}=0}\longrightarrow
   p^2 \;  \frac{\mu g^2}{2}
 \left( m_{\!\ssc A}^2  + |m|^2 \right)
I_{\!\ssc 34}(m_{\!\ssc A}^2,m_{\!\ssc A}^2)
\nonumber \\ && \qquad \qquad
  + \frac{\mu g^2}{2} \Big[ \left ( m_{\!\ssc A}^2  + |m|^2 \right)
  I_{\!\ssc 2\!F}(m_{\!\ssc A}^2,m_{\!\ssc A}^2) 
 -I_{\!\ssc F}(m_{\!\ssc A}^2) \Big]\;, 
\eea
\bea
\frac{-1}{2} \Sigma_{\!\ssc C\!C}
&=&
p^2\frac{g^2}{8} \Big\{
 \left[ m_{\!\ssc A_-}^2+(|m|-|\tilde{\eta}|)^2 \right]^2
              I_{\!\ssc 34}(m_{\!\ssc A_-}^2,m_{\!\ssc A_-}^2)
   -  2 |\tilde{\eta}| |m| m_{\!\ssc A_+}^2
		     I_{\!\ssc 34}(m_{\!\ssc A_-}^2,m_{\!\ssc A_+}^2)
\nonumber\\&&\qquad
    + 2 |\tilde{\eta}| |m|m_{\!\ssc A_-}^2 
         I_{\!\ssc 34}(m_{\!\ssc A_+}^2,m_{\!\ssc A_-}^2)
     +\left[ m_{\!\ssc A_+}^2+(|m|+|\tilde{\eta}|)^2 \right] ^2 
   I_{\!\ssc 34}(m_{\!\ssc A_+}^2,m_{\!\ssc A_+}^2)
\nonumber\\ &&\qquad
      -3 \! \left[ m_{\!\ssc A_-}^2+(|m|-|\tilde{\eta}|)^2 \right]
              \!    I_{\!\ssc 2\!F}(m_{\!\ssc A_-}^2,m_{\!\ssc A_-}^2)
   -3 \! \left[ m_{\!\ssc A_+}^2+(|m|+|\tilde{\eta}|)^2 \right]
        \!     I_{\!\ssc 2\!F}(m_{\!\ssc A_+}^2,m_{\!\ssc A_+}^2)
\nonumber\\&&\qquad\;\;		
    + 4 m_{\!\ssc A_-}^2 \left[ m_{\!\ssc A_-}^2+(|m|-|\tilde{\eta}|)^2 \right]
         I_{\!\ssc 3\!F}(m_{\!\ssc A_-}^2,m_{\!\ssc A_-}^2,m_{\!\ssc A_-}^2)
\nonumber\\&&\qquad\quad
  +4 m_{\!\ssc A_+}^2 \left[ m_{\!\ssc A_+}^2+(|m|+|\tilde{\eta}|)^2 \right] 
        I_{\!\ssc 3\!F}(m_{\!\ssc A_+}^2,m_{\!\ssc A_+}^2,m_{\!\ssc A_+}^2)
\Big\}
\nonumber\\&& \quad
+\frac{g^2}{16} \Big\{
16-8|m|^2I_{\!\ssc F}(|m|^2)
+\left[ m_{\!\ssc A_+}^2-5m_{\!\ssc A_-}^2-8(|m|-|\tilde{\eta}|)^2\right]
     I_{\!\ssc F}(m_{\!\ssc A_-}^2)
\nonumber\\&&\qquad\qquad
  -\left[ 5m_{\!\ssc A_+}^2-m_{\!\ssc A_-}^2+8(|m|+|\tilde{\eta}|)^2 \right]
     I_{\!\ssc F}(m_{\!\ssc A_+}^2)
\nonumber\\ &&\qquad\qquad 
  +\left[ 2m_{\!\ssc A_-}^2m_{\!\ssc A_+}^2
           -m_{\!\ssc A_+}^2m_{\!\ssc A_+}^2-m_{\!\ssc A_-}^2m_{\!\ssc A_-}^2 \right]
		    I_{\!\ssc 2\!F}(m_{\!\ssc A_-}^2,m_{\!\ssc A_+}^2)
\nonumber\\&&\qquad\qquad  
  +2  \left[  m_{\!\ssc A_-}^2+(|m|-|\tilde{\eta}|)^2 \right]^2
			  I_{\!\ssc 2\!F}(m_{\!\ssc A_-}^2,m_{\!\ssc A_-}^2)
\nonumber\\&&\qquad\qquad 
  +2 \left[ m_{\!\ssc A_+}^2+(|m|+|\tilde{\eta}|)^2 \right]^2
				I_{\!\ssc 2\!F}(m_{\!\ssc A_+}^2,m_{\!\ssc A_+}^2)
\Big\} + \cdots \;,
\nonumber \\ && _{\tilde{\eta}=0}\longrightarrow
p^2\; \frac{g^2}{4} \Big[
    4  m_{\!\ssc A}^2  \left( m_{\!\ssc A}^2+|m|^2\right) 
      I_{\!\ssc 3\!F}(m_{\!\ssc A}^2,m_{\!\ssc A}^2,m_{\!\ssc A}^2)
    + \left( m_{\!\ssc A}^2+|m|^2 \right)^2       I_{\!\ssc 34}(m_{\!\ssc A}^2,m_{\!\ssc A}^2)
\nonumber\\
&&\qquad\qquad\qquad
   - 3\left(m_{\!\ssc A}^2+|m|^2\right)I_{\!\ssc 2\!F}(m_{\!\ssc A}^2,m_{\!\ssc A}^2) \Big]
\nonumber\\
&&\qquad \qquad
+\frac{g^2}{4} \Big[ 4-2|m|^2I_{\!\ssc F}(|m|^2)
 -2\left(m_{\!\ssc A}^2+2|m|^2\right)I_{\!\ssc F}(m_{\!\ssc A}^2)
\nonumber\\
&&\qquad \qquad\qquad
 +\left(m_{\!\ssc A}^2+|m|^2\right)^2I_{\!\ssc 2\!F}(m_{\!\ssc A}^2,m_{\!\ssc A}^2)
\Big] \;,
\eea
\bea
-\Sigma_{\!\ssc N\!N^*}
&=&
%
 p^2 \; \frac{g^2}{2}
\Big\{  \left(|m|-|\tilde{\eta}|\right)^2
I_{\!\ssc 34}(m_{\!\ssc A_-}^2,m_{\!\ssc A_-}^2)
  + \left(|m|+|\tilde{\eta}|\right)^2
I_{\!\ssc 34}(m_{\!\ssc A_+}^2,m_{\!\ssc A_+}^2)
\nonumber \\ && \qquad\quad
 +  |m| \left(|m|+|\tilde{\eta}|\right)
I_{\!\ssc 34}(m_{\!\ssc A_-}^2,m_{\!\ssc A_+}^2)
 +  |m| \left(|m|-|\tilde{\eta}|\right)
I_{\!\ssc 34}(m_{\!\ssc A_+}^2,m_{\!\ssc A_-}^2)
\nonumber \\ && \quad
+ \frac{g^2}{2} \Big\{
  \left(|m|-|\tilde{\eta}|\right)^2   I_{\!\ssc 2\!F}(m_{\!\ssc A_-}^2,m_{\!\ssc A_-}^2)
+ \left(|m|+|\tilde{\eta}|\right)^2  I_{\!\ssc 2\!F}(m_{\!\ssc A_+}^2,m_{\!\ssc A_+}^2)
\nonumber \\ && \qquad\quad 
 + 2|m|^2 I_{\!\ssc 2\!F}(m_{\!\ssc A_-}^2,m_{\!\ssc A_+}^2) 
-  I_{\!\ssc F}(m_{\!\ssc A_-}^2) - I_{\!\ssc F}(m_{\!\ssc A_+}^2) \Big\}
  + \cdots \;,
\nonumber \\ && _{\tilde{\eta}=0}\longrightarrow
   p^2 \;  2{g^2} |m|^2
I_{\!\ssc 34}(m_{\!\ssc A}^2,m_{\!\ssc A}^2)
 +{g^2} \left[   2|m|^2 I_{\!\ssc 2\!F}(m_{\!\ssc A}^2,m_{\!\ssc A}^2) 
  - I_{\!\ssc F}(m_{\!\ssc A}^2) \right] \;.
\eea
There are more mixing terms which vanish with $\tilde{\eta}$, as follows:
\bea
-\Sigma_{\!\ssc N\!N}
&=&
 p^2 \; \frac{g^2}{4} \frac{\tilde{\eta}^2}{|\tilde{\eta}|^2} \Big[ 
\left(|m|-|\tilde{\eta}|\right)^2
I_{\!\ssc 34}(m_{\!\ssc A_-}^2,m_{\!\ssc A_-}^2)
 + \left(|m|+|\tilde{\eta}|\right)^2
I_{\!\ssc 34}(m_{\!\ssc A_+}^2,m_{\!\ssc A_+}^2)
\nonumber \\ && \qquad\qquad\quad
- |m| \left(|m|-|\tilde{\eta}|\right)
I_{\!\ssc 34}(m_{\!\ssc A_-}^2,m_{\!\ssc A_+}^2)
- |m| \left(|m|+|\tilde{\eta}|\right)
I_{\!\ssc 34}(m_{\!\ssc A_+}^2,m_{\!\ssc A_-}^2)
\Big] 
\nonumber \\ && \qquad
 + \frac{g^2}{4} \frac{\tilde{\eta}^2}{|\tilde{\eta}|^2} \Big[ 
 \left(|m|-|\tilde{\eta}|\right)^2  I_{\!\ssc 2\!F}(m_{\!\ssc A_-}^2,m_{\!\ssc A_-}^2)
 + \left(|m|+|\tilde{\eta}|\right)^2  I_{\!\ssc 2\!F}(m_{\!\ssc A_+}^2,m_{\!\ssc A_+}^2)
\nonumber \\ && \qquad\qquad\qquad
 - 2|m|^2  I_{\!\ssc 2\!F}(m_{\!\ssc A_-}^2,m_{\!\ssc A_+}^2)
\Big] 
\;,
\eea
\bea
-\Sigma_{\!\ssc C\!N^*}
&=&
p^2 \; \frac{g^2}{2} \frac{\tilde{\eta}}{|\tilde{\eta}|} \bigg\{
\left(|m|+|\tilde{\eta}|\right)
 \left[  m_{\!\ssc A_+}^2+   \left(|m|+|\tilde{\eta}|\right)^2 \right]
I_{\!\ssc 34}(m_{\!\ssc A_+}^2,m_{\!\ssc A_+}^2)
-m_{\!\ssc A_+}^2 |m|  I_{\!\ssc 34}(m_{\!\ssc A_-}^2,m_{\!\ssc A_+}^2)
\nonumber \\ && \qquad\quad
+ m_{\!\ssc A_-}^2 |m| I_{\!\ssc 34}(m_{\!\ssc A_+}^2,m_{\!\ssc A_-}^2)
 - \left(|m|-|\tilde{\eta}|\right)
 \left[  m_{\!\ssc A_-}^2+   \left(|m|-|\tilde{\eta}|\right)^2 \right]
I_{\!\ssc 34}(m_{\!\ssc A_-}^2,m_{\!\ssc A_-}^2)
\bigg\}
\nonumber \\ && \;\;
+ \frac{g^2}{2} \frac{\tilde{\eta}}{|\tilde{\eta}|} \Big\{
\left(3|m|-2|\tilde{\eta}|\right)  I_{\!\ssc F}(m_{\!\ssc A_-}^2)
- \left(3|m|+2|\tilde{\eta}|\right)  I_{\!\ssc F}(m_{\!\ssc A_+}^2)
-4 |m|^2|\tilde{\eta}| I_{\!\ssc 2\!F}(m_{\!\ssc A_-}^2,m_{\!\ssc A_+}^2)
\nonumber \\ && \qquad\qquad\qquad
 -\left(|m|-|\tilde{\eta}|\right)
\left[  m_{\!\ssc A_-}^2+   \left(|m|-|\tilde{\eta}|\right)^2 \right]
 I_{\!\ssc 2\!F}(m_{\!\ssc A_-}^2,m_{\!\ssc A_-}^2)
\nonumber \\ && \qquad\qquad\qquad
 + \left(|m|+|\tilde{\eta}|\right)
\left[  m_{\!\ssc A_+}^2+   \left(|m|+|\tilde{\eta}|\right)^2 \right]
I_{\!\ssc 2\!F}(m_{\!\ssc A_+}^2,m_{\!\ssc A_+}^2)
\Big\} \;,
\eea
\bea
-\Sigma_{\!\ssc D\!N^*}
&=&
p^2 \; \frac{\mu g^2}{2} \frac{\tilde{\eta}}{|\tilde{\eta}|} \Big[ 
  \left(|m|+|\tilde{\eta}|\right)
I_{\!\ssc 34}(m_{\!\ssc A_+}^2,m_{\!\ssc A_+}^2)
-\left(|m|-|\tilde{\eta}|\right)
I_{\!\ssc 34}(m_{\!\ssc A_-}^2,m_{\!\ssc A_-}^2)
\nonumber \\ && \qquad\qquad\quad
-|m|  I_{\!\ssc 34}(m_{\!\ssc A_-}^2,m_{\!\ssc A_+}^2)
+ |m| I_{\!\ssc 34}(m_{\!\ssc A_+}^2,m_{\!\ssc A_-}^2)
\Big]   
\nonumber \\ && \;
+ \frac{\mu g^2}{2} \frac{\tilde{\eta}}{|\tilde{\eta}|} \Big\{
 \left(|m|+|\tilde{\eta}|\right)  I_{\!\ssc 2\!F}(m_{\!\ssc A_+}^2,m_{\!\ssc A_+}^2)
 - \left(|m|-|\tilde{\eta}|\right) I_{\!\ssc 2\!F}(m_{\!\ssc A_-}^2,m_{\!\ssc A_-}^2)
\Big\} \;,
\eea
with also the complex conjugates for the last three, {\em i.e.}
$-\Sigma_{\!\ssc N^*\!N^*}$, $-\Sigma_{\!\ssc C\!N}$, and $-\Sigma_{\!\ssc D\!N}$.

It is somewhat of a surprise that all the scalar actually becomes kinetic, including 
$D$ and $N$. The latter are introduced as auxiliary components of mass dimension 
two. One should hence consider $\frac{D}{\mu}$ and $\frac{N}{\mu}$ instead.  For 
general case, the complex $\frac{N}{\mu}$ has to be expanded into the real 
components first. One has then to diagonalize the kinetic term matrix for all the 
real scalars  to find the proper wavefunction renormalization factors for the 
canonical modes, and subsequently diagonalize the mass-square matrix,
with the tree-level terms included, of the latter for the eigenvalues.

\section{Propagator expressions for the most general case}
We give here the superfield propagator expressions for the most general case,
{\em i.e.} all soft supersymmetry breaking parameters are included. The 
propagator(s) used in our model above is the case with the soft mass term
$-\frac{1}{2} \eta \theta^2 \Phi^2$ in the superpotential vanishing. The 
expressions have, apparently, not been explicitly given before, and may be 
useful in some future studies.

The free-field Lagrangian for a single chiral superfield
$\Phi=A+\sqrt{2} \psi \theta + F  \theta^2$  admitting all supersymmetric
and (soft) supersymmetry breaking  mass parameter can be written as
\bea
{\mathcal{L}}_o &=& \int\!\! d^4 \theta \; \bar{\Phi} \Phi
(1-\tilde{\eta}\theta^2
-\tilde{\eta}^* \bar{\theta}^2-\tilde{m}^2 \theta^2 \bar{\theta}^2)
+ \left[ \int\!\! d^2 \theta \;\frac{1}{2} (m-\eta \theta^2)  \Phi^2 \delta^2(\bar{\theta})
+ h.c. \right] .
\eea
Again, we allow a complex $m$. Soft supersymmetry breaking parameters $\tilde{\eta}$
and ${\eta}$ are also complex while the most familiar soft mass $\tilde{m}^2$ is
real. The superfield propagators are given by
\bea
&&
\langle T(\Phi(1) \Phi^\dagger(2)) \rangle =
\frac{-i}{p^2+|m|^2} \delta^4_{\ssc 12}
- \frac{ i [\tilde\eta (Q-2|m|^2)+m^*\eta]}{Q^2- |\eta-2m \tilde{\eta}|^2 }
  \, {\theta_{\!\ssc 1}}^2 \delta^4_{\ssc 12}
- \frac{ i [\tilde\eta^* (Q-2|m|^2)+m\eta^*]}{Q^2- |\eta-2m \tilde{\eta}|^2 }
  \, \bar{\theta_{\!\ssc 1}}^2 \delta^4_{\ssc 12}
\nonumber \\ && \qquad
+ i\frac{(-p^2|\tilde\eta|^2+ \tilde{m}^2|m|^2)Q + 4p^2 |m|^2|\tilde\eta|^2 -(p^2-|m|^2)
  (m^* \eta \tilde{\eta}^*+m \eta^*\tilde{\eta})-|m|^2 |\eta|^2}
    {(p^2+|m|^2|)(Q^2- |\eta-2m \tilde{\eta}|^2) }
   \theta_{\!\ssc 1}^2 \bar{\theta_{\!\ssc 1}}^2 \delta^4_{\ssc 12}
\nonumber \\ && \qquad
+  i\frac{(\tilde{m}^2 + |\tilde\eta|^2)Q - |\eta-2m \tilde{\eta}|^2}
  {(p^2+|m|^2)(Q^2- |\eta-2m \tilde{\eta}|^2) }
  \left[ \frac{D_{\!\ssc 1}^2 \theta_{\!\ssc 1}^2 \bar{\theta_{\!\ssc 1}}^2 \overline{D}_{\!\ssc 1}^2}
    {16} \right] \delta^4_{\ssc 12}\;,
\eea
and
\bea
&&
\langle T(\Phi(1) \Phi(2)) \rangle =
\frac{i \, {m}^*}{p^2(p^2+|m|^2)} \frac{D_{\!\ssc 1}^2}{4} \delta^4_{\ssc 12}
-\frac{i(\eta^*-2m^* \tilde{\eta}^*)}{Q^2- |\eta-2m \tilde{\eta}|^2 }
   \frac{ D_{\!\ssc 1}^2  \bar{\theta_{\!\ssc 1}}^2}{4}  \delta^4_{\ssc 12}
\nonumber  \\&& \qquad\qquad 
+  i \, \frac{2m^* \tilde\eta (p^2+\tilde m^2 ) + m^{*2} \eta + \eta^* \tilde\eta^2 }
      {Q^2- |\eta-2m \tilde{\eta}|^2 }
 \frac{D_{\!\ssc 1}^2 {\theta_{\!\ssc 1}}^2}{4p^2} \delta^4_{\ssc 12}
%
\nonumber  \\&& \quad
+ i\, \frac{  m^*[(\tilde{m}^2 + |\tilde\eta|^2)Q - |\eta-2m \tilde{\eta}|^2]
       -\tilde\eta (\eta^*- 2m^*\tilde\eta^*)(p^2+|m|^2)}
      {(p^2+|m|^2)(Q^2- |\eta-2m \tilde{\eta}|^2) } \!
     \left[ \frac{ D_{\!\ssc 1}^2 \theta_{\!\ssc 1}^2 \bar{\theta_{\!\ssc 1}}^2  }{4}
      \!+\! \frac{\bar{\theta_{\!\ssc 1}}^2 \theta_{\!\ssc 1}^2 D_{\!\ssc 1}^2}{4} \right]
         \!\!\delta^4_{\ssc 12} \, .
\nonumber \\
\eea
where $Q=p^2+ |m|^2 + \tilde{m}^2+ |\tilde\eta|^2$.
The corresponding component field propagators are given by
\bea
\langle T(A\,A^*)\rangle
&=&\frac{-i ( p^2+ |m|^2 + \tilde{m}^2+ |\tilde\eta|^2 )}
   { ( p^2+ |m|^2 + \tilde{m}^2+ |\tilde\eta|^2)^2 - |\eta-2m \tilde{\eta}|^2 }  \;,
\nonumber\\
\langle T(A\,A)\rangle
&=&\frac{ i (\eta^*-2m^* \tilde{\eta}^*) }
   { ( p^2+ |m|^2 + \tilde{m}^2+ |\tilde\eta|^2)^2 - |\eta-2m \tilde{\eta}|^2  }  \;,
\nonumber\\
\langle T(F\,F^*)\rangle
&=&\frac{i [(p^2  + \tilde{m}^2) ( p^2+ |m|^2 + \tilde{m}^2+ |\tilde\eta|^2)-|\eta-m \tilde{\eta}|^2
+|m \tilde{\eta}|^2 ]}
   { ( p^2+ |m|^2 + \tilde{m}^2+ |\tilde\eta|^2)^2 - |\eta-2m \tilde{\eta}|^2 }    \;,
\nonumber\\
\langle T(F\,F)\rangle
&=&\frac{ i [ 2 m^* \tilde{\eta}( p^2+  \tilde{m}^2)
+ m^{*2}\eta + \eta^* \tilde{\eta}^2 ]}
   { ( p^2+ |m|^2 + \tilde{m}^2+ |\tilde\eta|^2)^2 - |\eta-2m \tilde{\eta}|^2  }    \;,
\nonumber\\
\langle T(A\,F)\rangle
&=&\frac{i [m^*   ( p^2+ |m|^2 + \tilde{m}^2- |\tilde\eta|^2)+\eta^*\tilde\eta]}
   { ( p^2+ |m|^2 + \tilde{m}^2+ |\tilde\eta|^2)^2 - |\eta-2m \tilde{\eta}|^2}   \;,
\nonumber\\
\langle T(AF^*)\rangle
&=&\frac{-i [\tilde\eta^* ( p^2- |m|^2 + \tilde{m}^2+ |\tilde\eta|^2) + m\eta^* ]}
   { ( p^2+ |m|^2 + \tilde{m}^2+ |\tilde\eta|^2)^2 - |\eta-2m \tilde{\eta}|^2 }   \;,
\nonumber\\
\langle T(\psi_{{\!\ssc R}_{\alpha}} \bar\psi_{{\!\ssc R}_{\dot\beta}})\rangle
&=& \frac {- i p_{\mu}\,\sigma^{\mu}_{\alpha \dot \beta}}  {  p^2+|m|^2 }    \;,
\nonumber\\
\langle T(\psi_{{\!\ssc R}_{\alpha}} \psi_{\!\ssc R}^{\beta})\rangle
&=& \frac {-i m^* \delta_\alpha^\beta } {  p^2+|m|^2 } \;.
\eea

Note that  the Lagrangian without all the masses has a $U(1)$ and
a $U(1)_{\!\ssc R}$ symmetry to which $\Phi$ carries both charges
(of 1).  $U(1)_{\!\ssc R}$ charges for the components $A$ and $F$
are 1 and -1, with $\psi$ neutral. We can assign corresponding
charges to the mass parameters and use them to help trace
and check the role of the parameters in the component
field propagators and the corresponding terms of the
superfield propagators.

\acknowledgments
We thank Dong-Won Jung for useful discussions. Questions and comments
also from anomalous referees for our companion short letter \cite{062}
prompted us to elaborate further on some discussion we presented
in an earlier draft which is mostly how Sec.VI came by. Thanks to all that
which help to improve the presentation in the current paper.
Y.C.,Y.-M.D., and O.K. are partially supported by research grant
NSC 102-2112-M-008-007-MY3, and Y.C. was further
supported by grants  NSC 103-2811-M-008-018 of the MOST
of Taiwan. G.F. is supported by research grant NTU-ERP-102R7701.

\end{document}